\documentclass[11pt]{article}
\usepackage{enumitem}
\usepackage[final]{acl}

\usepackage{siunitx}

\usepackage{subcaption}   %

\usepackage{tabularray}
\usepackage{times}
\usepackage{latexsym}
\usepackage{colortbl}
\usepackage{pifont}
\usepackage{bbding}
\usepackage{amsmath}

\usepackage[table]{xcolor}
\usepackage{soul}
\usepackage{pbox}
\usepackage[T1]{fontenc}

\usepackage[utf8]{inputenc}
\usepackage{tikz}
\usetikzlibrary{calc}  %
\usepackage{longtable}
\usepackage{listings}
\definecolor{myred}{RGB}{178, 34, 34} %
\definecolor{mygreen}{RGB}{34,139,34}   %
\definecolor{myred2}{RGB}{237, 211, 210} %
\definecolor{mygreen2}{RGB}{198, 232, 206} %
\definecolor{myblue2}{RGB}{218,232,252}

\definecolor{codegreen}{rgb}{0,0.6,0}
\definecolor{codegray}{rgb}{0.5,0.5,0.5}
\definecolor{codepink}{RGB}{252, 142, 172}
\definecolor{codepurple}{rgb}{0.58,0,0.82}
\definecolor{backcolour}{RGB}{245,245,245}

\definecolor{AuroraBlue}{HTML}{EAF3FB}
\definecolor{SkyBlue}{HTML}{BDE2F6}
\definecolor{SerenityBlue}{HTML}{73A5C6}
\definecolor{OxfordBlue}{HTML}{3373C4}
\definecolor{MidnightBlue}{HTML}{1E3F66}

\definecolor{LightGreen}{rgb}{0.88,1,0.88}
\definecolor{LightRed}{rgb}{1,0.88,0.88}
\newcommand{\greenhl}[1]{\sethlcolor{LightGreen}\hl{#1}}
\newcommand{\redhl}[1]{\sethlcolor{LightRed}\hl{#1}}

\definecolor{delim}{RGB}{20,105,176}
\definecolor{numb}{RGB}{106, 109, 32}
\definecolor{string}{rgb}{0.64,0.08,0.08}

\lstdefinestyle{jsonstyle}{
    showspaces=false,
    showtabs=false,
    breaklines=true,
    postbreak=\raisebox{0ex}[0ex][0ex]{\ensuremath{\color{gray}\hookrightarrow\space}},
    breakatwhitespace=true,
    basicstyle=\ttfamily\small,
    upquote=true,
    stringstyle=\color{string},
    literate=
     *{0}{{{\color{numb}0}}}{1}
      {1}{{{\color{numb}1}}}{1}
      {2}{{{\color{numb}2}}}{1}
      {3}{{{\color{numb}3}}}{1}
      {4}{{{\color{numb}4}}}{1}
      {5}{{{\color{numb}5}}}{1}
      {6}{{{\color{numb}6}}}{1}
      {7}{{{\color{numb}7}}}{1}
      {8}{{{\color{numb}8}}}{1}
      {9}{{{\color{numb}9}}}{1}
      {\{}{{{\color{delim}{\{}}}}{1}
      {\}}{{{\color{delim}{\}}}}}{1}
      {[}{{{\color{delim}{[}}}}{1}
      {]}{{{\color{delim}{]}}}}{1},
}

\lstdefinelanguage{json}{
    morestring=[b]", %
}

\lstdefinestyle{mystyle}{
    backgroundcolor=\color{backcolour},   
    commentstyle=\color{magenta},
    keywordstyle=\color{blue},
    numberstyle=\tiny\color{codegray},
    stringstyle=\color{codepurple},
    basicstyle=\fontfamily{\ttdefault}\footnotesize,
    breakatwhitespace=false,         
    breaklines=true,                 
    keepspaces=true,    
    frame=single,
    numbersep=5pt,                  
    showspaces=false,                
    showstringspaces=false,
    showtabs=false,                  
    tabsize=2,
    classoffset=1, %
    keywordstyle=\color{violet},
    classoffset=0,
}
\lstdefinelanguage{JavaScript}{
  keywords={typeof, new, true, false, catch, function, return, null, catch, switch, var, if, in, while, do, else, case, break},
  keywordstyle=\color{blue}\bfseries,
  ndkeywords={class, export, boolean, throw, implements, import, this},
  ndkeywordstyle=\color{darkgray}\bfseries,
  identifierstyle=\color{black},
  sensitive=false,
  comment=[l]{//},
  morecomment=[s]{/*}{*/},
  commentstyle=\color{purple}\ttfamily,
  stringstyle=\color{red}\ttfamily,
  morestring=[b]',
  morestring=[b]"
}

\usepackage{microtype}

\usepackage{inconsolata}

\usepackage{graphicx}

\usepackage{graphicx}
\usepackage{caption}
\usepackage{amssymb}
\usepackage{amsmath}
\usepackage{multirow}
\usepackage{multicol}
\usepackage{booktabs}
\usepackage{float}
\usepackage{placeins}
\usepackage{xcolor}
\usepackage{colortbl}
\definecolor{myred}{RGB}{237, 211, 210} %
\definecolor{mygreen}{RGB}{198, 232, 206} %
\definecolor{myblue}{RGB}{218,232,252}

\definecolor{myred}{RGB}{255,90,90}
\definecolor{mypink}{RGB}{239,43,159}
\definecolor{myupdate}{RGB}{254,243,222}
\definecolor{myfrozen}{RGB}{237,255,255}
\definecolor{ired}{RGB}{229,72,72}
\definecolor{igreen}{RGB}{80,219,144}
\definecolor{ired}{RGB}{247,142,142}
\definecolor{bluei}{RGB}{218,232,252}

\title{UniDataBench: Evaluating Data Analytics Agents Across Structured and Unstructured Data}

\author{
    Han Weng\textsuperscript{1}\thanks{Equal contribution.},
    Zhou Liu\textsuperscript{1,2}\footnotemark[1],
    Yuanfeng Song\textsuperscript{1}\thanks{Corresponding authors.},\\ %
    \textbf{Xiaoming Yin\textsuperscript{1}\footnotemark[2] ,
    Xing Chen\textsuperscript{1},
    Wentao Zhang\textsuperscript{2}}
    \\
    \textsuperscript{1}ByteDance, China  \\ \textsuperscript{2}Peking University, China
}

\begin{document}
\maketitle
\begin{abstract}
In the real business world, data is stored in a variety of sources, including structured relational databases, unstructured databases (e.g., NoSQL databases), or even CSV/excel files. The ability to extract reasonable insights across these diverse source is vital for business success. Existing benchmarks, however, are limited in assessing agents' capabilities across these diverse data types. To address this gap, we introduce \textbf{UniDataBench}, a comprehensive benchmark designed to evaluate the performance of data analytics agents in handling diverse data sources. 
Specifically, UniDataBench is originating from real-life industry analysis report and we then propose a pipeline to remove the privacy and sensitive information. It encompasses a wide array of datasets, including relational databases, CSV files to NoSQL data, reflecting real-world business scenarios, and provides unified framework to assess how effectively agents can explore multiple data formats, extract valuable insights, and generate meaningful summaries and recommendations. 
Based on UniDataBench, we propose a novel LLM-based agent named \textbf{ReActInsight}, an autonomous agent that performs end-to-end analysis over diverse data sources by automatically discovering cross-source linkages, decomposing goals, and generating robust, self-correcting code to extract actionable insights. Our benchmark and agent together provide a powerful framework for advancing the capabilities of data analytics agents in real-world applications.

\end{abstract}

\section{Introduction}

The ultimate goal of modern enterprise analytics is to support effective decision-making by distilling vast amounts of raw data into insights \cite{10.1145/2532780.2544424, black2023business, ghazal2013bigbench, sharma2014transforming}. This process typically begins with a high-level business objective and proceeds through an iterative exploration of data. In contemporary business intelligence, however, the most valuable insights are seldom found within a single, well-structured dataset. Instead, they emerge from the complex synthesis of a fragmented information landscape with diverse formats \cite{egg2025dabstepdataagentbenchmark, hu2024infiagentdabench}.

The rapid development of large language models (LLMs) has led to the emergence of data science agents capable of automating the deconstruction of complex problems and exploring insights within data by coordinating specialized tools \cite{ma-etal-2023-insightpilot, li2023sheetcopilot, zhang2024datacopilot, sahu2024insightbench, hong-etal-2025-data, bai2025insight, perez2025llm}. Despite extensive research in data science, current benchmarks fall short in measuring data agents' ability to perform comprehensive analysis across diverse data formats (e.g., databases, NoSQL, and text files) rather than just CSV files. Existing data science benchmarks such as Text2Analysis \cite{Text2Analysis2023aaai}, DABench \cite{hu2024infiagentdabench}, and DAStep \cite{egg2025dabstepdataagentbenchmark} focus on data analysis for specific problems. Furthermore, current multi-step data analysis benchmarks that start from high-level goals only focus on structured tabular data within a single enterprise management domain, with human-designed analytical conclusions further hindering their accuracy in measuring agents' capabilities in real-world business analysis scenarios \cite{sahu2024insightbench}.

To address these challenges, we propose UniDataBench, a comprehensive data analysis benchmark tailored for assessing the performance of data agents across various data formats.
This benchmark includes 100 datasets covering five business analysis scenarios, featuring diverse and varied input data file formats and difficulties. The tasks require agents to perform multi-step problem decomposition starting from a high-level goal, and to connect, query, and understand information from structured, semi-structured databases, and unstructured documents to generate insights. To ensure the value of the analytical problems and insights in the benchmark, we enlisted expert annotators to extract questions and corresponding relevant insights from real-world enterprise business data analysis reports, such as tracking product usage metrics, analyzing event-driven sales fluctuations, and monitoring changes in user feedback sentiment. Subsequently, we conducted data synthesis and visualization validation based on LLMs and manual inspection to ensure that the insights were indeed discoverable.

To bridge this gap, we further propose ReActInsight, an innovative autonomous agent engineered for end-to-end analysis across diverse data sources. ReActInsight initiates its workflow with a novel multi-source data exploration and linkage discovery phase. It automatically constructs a unified metadata representation (i.e., \textit{MetaGraph}) for all available data, then performs similarity analysis to discover potential join keys, formulating them into an actionable Join-Hint. Guided by this hint, the agent employs a hierarchical planning mechanism to decompose a high-level user goal into a series of answerable sub-questions. For each sub-question, it generates executable code, leveraging a robust self-correction and debugging module to ensure reliability and adaptive visualization techniques to uncover underlying patterns. Finally, ReActInsight synthesizes the results from all steps into a coherent summary, presenting the user with key insights and the detailed analytical process.

In summary, our contributions are as follows:
\begin{itemize}[leftmargin=*,itemsep=2pt,topsep=0pt,parsep=0pt]
\item We propose UniDataBench, a benchmark designed to evaluate the capability of agents in conducting data analysis starting from high-level goal under multiple data formats.

\item We propose ReActInsight, an agent extracts insights from multiple data formats through multi-source exploration, hierarchical planning, adaptive visualization. It achieves superior performance compared to other data analysis agents.

\item
We conduct extensive experiments on our benchmark, demonstrating the challenges of these tasks and validating the effectiveness of our proposed agent in tackling them.
\end{itemize}

 \begin{figure*}[t!]
    \centering
    \includegraphics[width = 1\textwidth]{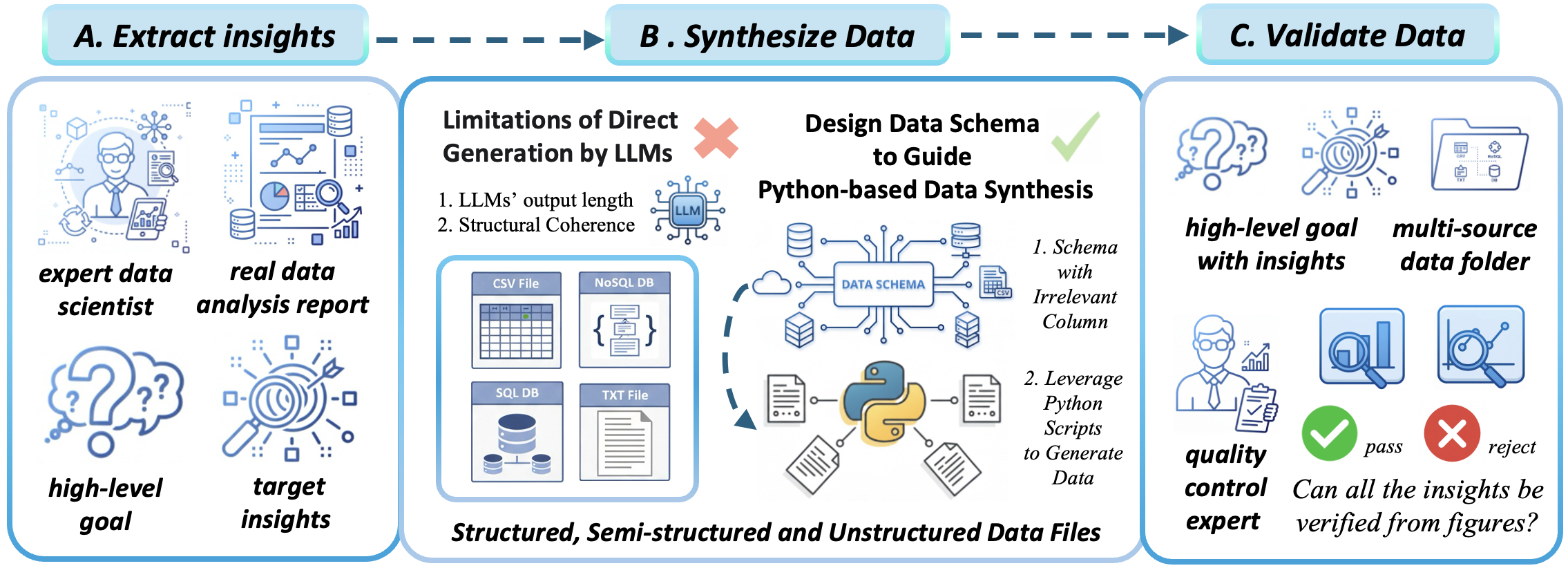}
    \vspace{-20pt}
    \caption{
    The three-stage generation pipeline for UniDataBench: (A) Extracting insights from real reports, (B) Designing a schema to guide Python-based data synthesis, and (C) Human validation via visualization.}
\label{fig:caption_figure_pipeline}
    \vspace{-15pt}
\end{figure*}

\section{Related Work}
\label{sec:bibtex}

\subsection{Data Science Benchmarks}
The evaluation of LLMs in data science has produced a rich landscape of benchmarks, initially focused on generating code for specific, isolated steps of the analytical workflow \cite{yu2018spider, 2023DS-1000,zhang-etal-2024-benchmarking-data,song2024marrying, qin-etal-2025-multitend, lu2025bridging}. 

For example, benchmarks like DS-1000 \cite{2023DS-1000} test the fine-grained code generation abilities of models on discrete data manipulation tasks using popular libraries.
As LLMs evolved and were augmented by agentic frameworks, it became clear that evaluating isolated skills was insufficient for gauging performance in real-world applications. This led to a new generation of benchmarks designed to evaluate an agent's ability to solve complex, end-to-end tasks by interacting with an execution environment. These benchmarks target a broader spectrum of data science activities \cite{huang2024code,jing2025dsbench,hu2024infiagentdabench,lei2024spider,egg2025dabstepdataagentbenchmark,sahu2024insightbench}. 
For example, InsightBench \cite{sahu2024insightbench} assesses the entire process of planning, coding, and insight generation, pushing the evaluation towards higher-level, multi-step reasoning.

Despite these advances, a critical limitation pervades the current landscape: these benchmarks are designed to assess capabilities within siloed data environments. Each task typically operates on a single, self-contained data source. This leaves a crucial gap in evaluating an agent's ability to perform analysis that requires synthesizing information from multiple data formats, which is a common requirement in modern business intelligence.

\subsection{Data Analytics Agents} 
The proliferation of LLMs has promoted the development of numerous agents designed to automate complex tasks. In the domain of data science, a variety of specialized agents have emerged to streamline analytical workflows \cite{li2023sheetcopilot,yao2023react,majumder2024discoverybench,ma-etal-2023-insightpilot,zhang2024datacopilot,sahu2024insightbench,hong-etal-2025-data}. 
For instance, SheetCopilot \cite{li2023sheetcopilot} is designed as a spreadsheet agent that can generate and execute a plan of operations on tabular data to achieve a desired outcome. Similarly, Data Copilot \cite{zhang2024datacopilot} can automatically perform data processing, prediction, and visualization based on questions. Other agents focus on the high-level goal of insight discovery. For example, 
AgentPoirot \cite{sahu2024insightbench} performs end-to-end data analysis to extract descriptive, diagnostic, predictive, and prescriptive insights. Concurrently, systems like Data Interpreter \cite{hong-etal-2025-data} enhance problem-solving capabilities in data science through advanced methods like dynamic planning and logical consistency checking.

While some prominent agents, such as OpenAI's Code Interpreter \cite{openai_code_interpreter}, Pandas Agent \cite{langchain2024pandas} and the ReAct-style Agent of DABStep \cite{egg2025dabstepdataagentbenchmark}, demonstrate the ability to process multiple data formats and answer questions about them, their operational paradigm is typically limited to addressing these formats in isolation. 
Our agent is specifically designed to focus on exploring data and discovering insights that span across both structured and unstructured data sources, thereby enriching the current landscape of agents in the field of data analytics by addressing the need for integrated, multi-source reasoning.

\begin{table*}[th!]
\centering
\scalebox{0.8}{
\begin{tabular}{l|lcccc}
\toprule
\rowcolor{gray!20} 
\textbf{Dataset} & \textbf{Analysis Source} & \textbf{Multi-source Support} & \textbf{Task Size} & \textbf{File Size} & \textbf{Rows/File} \\
\midrule
\multicolumn{6}{l}{\textit{Specific Data Analysis - A specific answer provided for Question: What are the applicable fee IDs for Rafa\_AI in 2023?}} \\
\midrule
DS-1000 \cite{2023DS-1000} & StackOverflow & \textcolor{red}{\XSolidBrush} (data.frame) & 1000 & - & - \\
DSEval \cite{zhang-etal-2024-benchmarking-data} & LLM Generation \& Human & \textcolor{red}{\XSolidBrush} (csv) & 825 & 44 &  \cellcolor{OxfordBlue!30} 23630 \\
Text2Analysis \cite{Text2Analysis2023aaai} & LLM Generation \& Human & \textcolor{red}{\XSolidBrush} (csv) & 2249 & 347 & - \\
DABench \cite{hu2024infiagentdabench} & LLM Generation \& Human  & \textcolor{red}{\XSolidBrush} (csv) & 257 & 52 & \cellcolor{OxfordBlue!10}2024 \\
DABStep \cite{egg2025dabstepdataagentbenchmark} & \cellcolor{green!45!yellow!21} Real-world Data Analysis & \textcolor{green!50!black}{\checkmark} (3 types) & 450 & 7 & \cellcolor{OxfordBlue!20} 3121 \\
\midrule
\multicolumn{6}{l}{\textit{High Level Data Analysis - Insights are extracted for Goal: Find the discrepancy and imbalance in incidents assigned}} \\
\midrule
InsightBench \cite{sahu2024insightbench}  & LLM \& Human Planting & \textcolor{red}{\XSolidBrush} (csv) & 100 & 112 & \cellcolor{AuroraBlue!100} 545 \\
\textbf{UniDataBench (Ours)} & \cellcolor{green!45!yellow!21} Real-world Data Insights & \textcolor{green!50!black}{\checkmark} (4 types) & 100 & 223 & \cellcolor{OxfordBlue!30} 17945 \\
\bottomrule
\end{tabular}
}
\vspace{-5pt}
\caption{Comparison of UniDataBench with other benchmarks. UniDataBench is distinguished by its emphasis on high-level data analysis, which extends beyond the limitations of CSV-only data to support multiple file formats.}
\vspace{-15pt}
\label{Table:DatasetTable}
\end{table*}

\section{The proposed UniDataBench}

\subsection{Overview}
To address a key limitation of existing high-level data analysis benchmarks, which are constrained to a structured data format (e.g. CSV or MySQL), we propose UniDataBench. Our benchmark is built around insights collected directly from real-world enterprise data analysis reports. Deriving these insights requires synthesizing information from diverse data formats, including structured (CSV, SQLite), semi-structured (NoSQL), and unstructured (TXT). 
In this sections, we introduce the dataset construction pipeline illustrated in Figure~\ref{fig:caption_figure_pipeline}. 

\subsubsection{Extract insights from Real Reports}
\label{sec:data-schema-generation}
Unlike insights that are synthetically generated by LLMs or artificially constructed from human-implanted trends~\cite{zhang-etal-2024-benchmarking-data,Text2Analysis2023aaai,hu2024infiagentdabench,sahu2024insightbench}, the insights in UniDataBench stem from real-world business reports analyzing areas like product metrics, sales figures, and market competition. These insights are then carefully classified, filtered, combined, and mapped to high-level analytical goals.

\paragraph{Types of Insights} The insights extracted can be classified into four types, each common in data analysis: (i) \textit{Trend}: identify patterns or directional movements in data over time or sequence. (ii) \textit{Comparison}: involve segmenting data to compare the performance or behavior of different groups. (iii) \textit{Extreme Value}: focus on pinpointing maximum or minimum values to identify outliers or key performers. (iv) \textit{Attribution}: aim to uncover causal or correlational relationships between variables to explain why something is happening.

\subsubsection{Data Synthesis and Validation}
\label{sec:data-synthesis}

Though we obtain real-life insights via the above-mentioned step, we often lack the most original underlying data related to these insights. This is because such underlying data are usually sensitive, and current business reports only release conclusive insights. To address this issue, this step focuses on automatically constructing underlying data that aligns with the trends of these insights.

As illustrated in Figure~\ref{fig:caption_figure_pipeline}, direct synthesis of large-scale, complex data by LLMs is constrained by their inherent length limitations and the challenge of maintaining long-range structural coherence \cite{tang-etal-2024-struc}. To address this, we adopt a three-step approach for data generation. First, leveraging the extracted high-level questions and corresponding insights, we construct comprehensive instruction prompts for LLMs by incorporating scenario-specific context (e.g., domain formulas such as \textit{Lead Rate = CTR * CVR}). These prompts guide LLMs to design multi-source data schemas, whose outputs then undergo manual inspection to ensure coherence and correctness. Second, we proactively introduce distractor columns that are irrelevant to the analytical objectives but related to scenario to increase analytical complexity and better mimic real-world scenarios. Finally, we instruct the LLM to generate a Python script for data synthesis with former information. The generation of unstructured data follows a separate process. We use the required analytical information as a core basis, prompting an LLM to expand upon it and generate a realistic, simulated enterprise document. A detailed example of data synthesis and validation is provided in  Appendix~\ref{sec:appendix_synth_example}.

During the data validation process, we utilized Python scripts to generate visualizations related to insights. These charts are then subjected to a human-in-the-loop review by data analysis experts. Their task is to confirm that every related insight is clearly discoverable from the visual evidence.

\begin{table}[t!]
\centering
\vspace{5pt}
\label{tab:dataset_stats_compact}
\resizebox{\linewidth}{!}{
\begin{tabular}{@{} l r @{\hskip 3em} l r @{}}  %
    \toprule
    \multicolumn{2}{c}{\textbf{Insights (Size: 439)}} & \multicolumn{2}{c}{\textbf{Files (Size: 223)}} \\  %
    \cmidrule(r){1-2} \cmidrule(l){3-4}  %
    \textbf{Category} & \textbf{Size (\%)} & \textbf{Type} & \textbf{Size (\%)} \\  %
    \midrule
    Trend  & 112 (25.5\%) & CSV  &  110 (49.3\%) \\
    Extreme  &  126 (28.7\%) & DB&  49 (22.0\%) \\
    Comparison  &  110 (25.1\%)& NoSQL &  32 (14.3\%) \\
    Attribution &  91 (20.7\%) & TXT  &  32 (14.3\%) \\
    \bottomrule
\end{tabular}
}
\vspace{-5pt}
\caption{Insights and File Types in UniDataBench.}
\label{Table:Dataset_Statistics}
\vspace{-10pt}
\end{table}

\begin{figure}[t!]
    \centering
    \includegraphics[width = 0.35\textwidth]
    {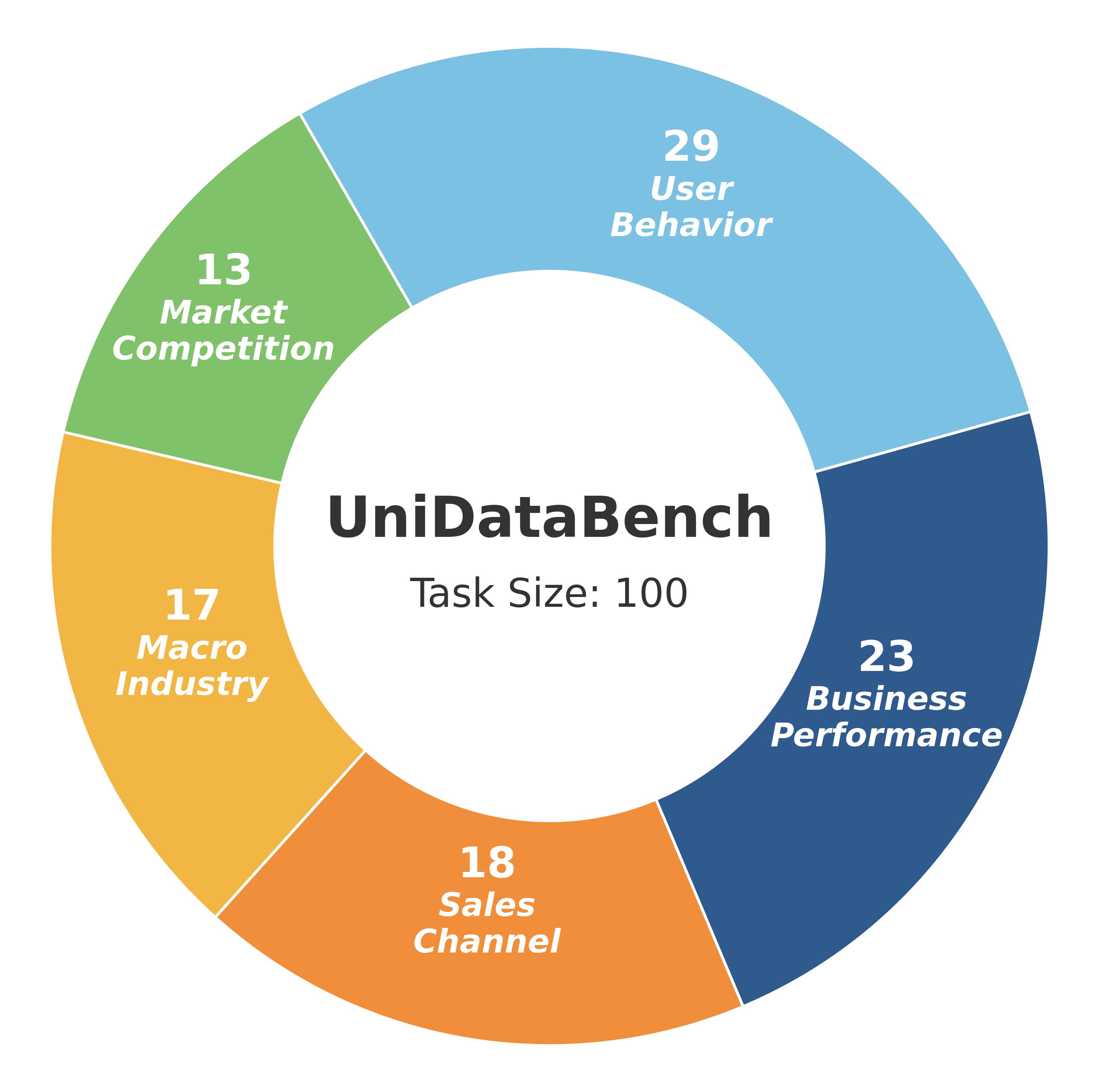}
    \vspace{-5pt}
    \caption{Domain Distribution in UniDataBench} 
    \label{fig:domain_dist}
    \vspace{-15pt}
\end{figure}

\subsection{Statistics of {UniDataBench}}
UniDataBench is a comprehensive benchmark for high-level data analysis. It features 100 analytical tasks designed to elicit 439 target insights from a corpus of 223 data files. The tasks are stratified into three difficulty levels based on the number of data formats required for analysis: 25 easy (single format), 50 medium (two formats), and 25 hard (three or four formats). As illustrated in Figure~\ref{fig:domain_dist}, the tasks span five core business domains, while Table~\ref{Table:Dataset_Statistics} provides a detailed breakdown of the insight categories and file type distributions. More comprehensive statistics are provided in Appendix~\ref{sec:appendix_data_statistics}.

\begin{figure*}[t!]
    \centering
    \includegraphics[width=1.0\textwidth]{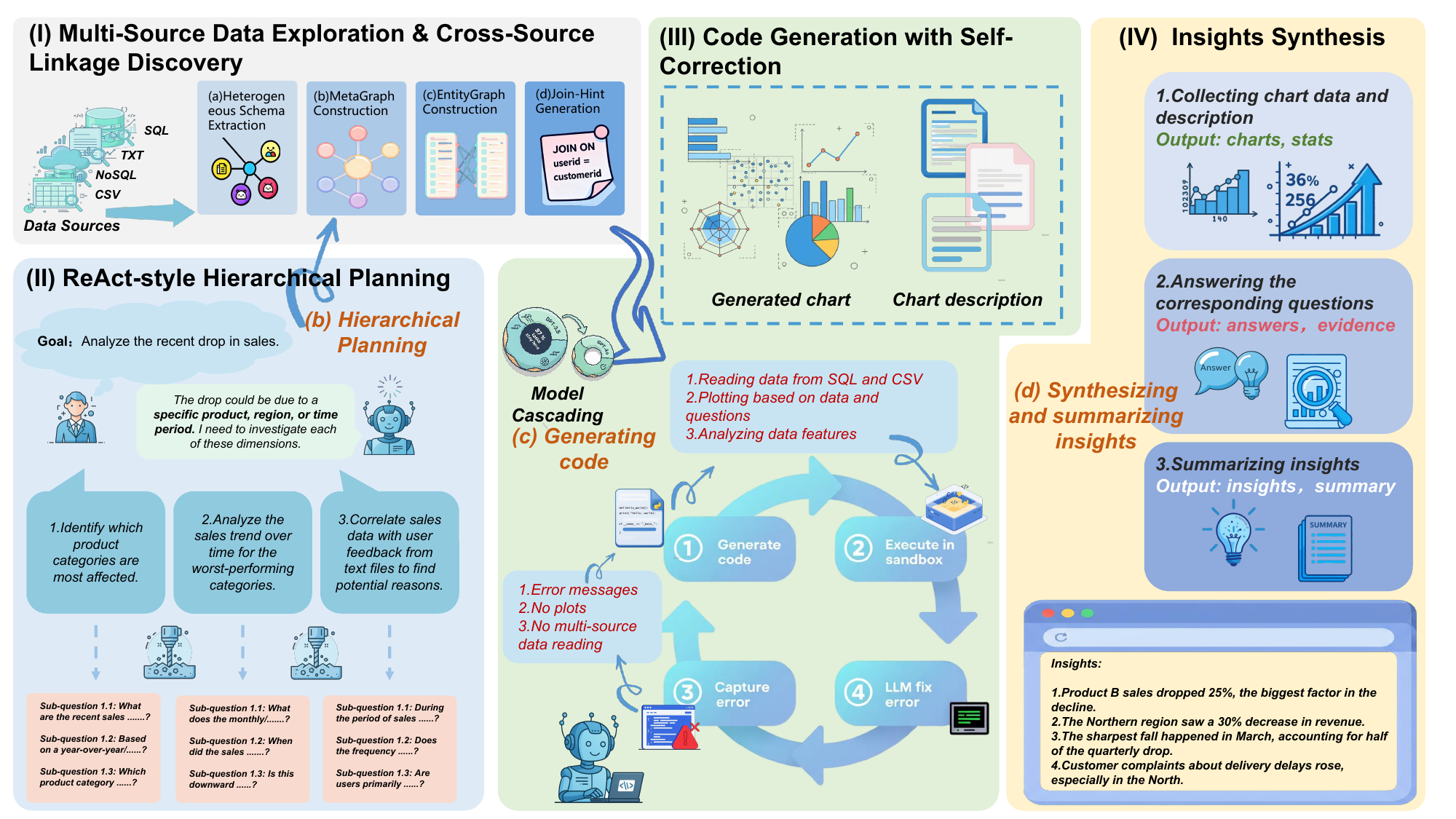}
    \vspace{-20pt}
    \caption{The workflow starts with (I) multi-source data exploration and cross-source linkage discovery, proceeds to (II) ReAct-style hierarchical planning that decomposes the analytical goal into sub-questions, continues with (III) automatic code generation augmented by an iterative self-correction loop, and culminates in (IV) insight synthesis that distills visual evidence and answers into conclusions.}
    \vspace{-10pt}
    \label{fig:Architecture}
\end{figure*}

\section{The Proposed ReActInsight}

\subsection{Motivation and Overview}
Despite their proficiency in executing siloed operations on single data modalities, exisiting agents usually lack the capability to orchestrate complex analytical workflows and synthesize coherent insights from diverse data formats. To alleivate this limitation, we introduce \textbf{ReActInsight}, a novel autonomous agent designed for end-to-end data analysis across both structured and unstructured data sources. ReActInsight moves beyond simple reactive execution by implementing a sophisticated, multi-layered reasoning process inspired by the ReAct framework. 

As shown in Figure~\ref{fig:Architecture}, its workflow begins with an efficient \textbf{multi-source data exploration} phase to gain a preliminary understanding of the vast and varied data landscape. Following this, ReActInsight employs a \textbf{hierarchical planning mechanism}, decomposing the primary analytical goal into a series of manageable sub-goals and specific, answerable sub-questions. For each sub-question, the agent autonomously generates executable code, leveraging \textbf{diverse and adaptive visualization} techniques to uncover underlying patterns. Crucially, it is equipped with a robust \textbf{self-correction and debugging} module to ensure reliability. 
Finally, ReActInsight culminates its analysis by presenting the user with the discovered insights, a conclusive summary, and the detailed analytical steps that led to these findings.

\subsection{Architecture of ReActInsight}
The architecture of ReActInsight is designed as a multi-stage, iterative pipeline that enables comprehensive data analysis from initial exploration to final insights. Each component is engineered to handle the complexity and heterogeneity of real-world data environments.

\subsubsection{Multi-Source Data Exploration and Cross-Source Linkage Discovery}
To overcome the limitations of analyzing data sources in isolation, ReActInsight initiates its process with an intelligent exploration and linkage discovery pipeline. This phase is critical for building a semantic understanding of how disparate datasets relate to one another. The process is broken down into four key steps:
\begin{enumerate}
[leftmargin=12pt, itemsep=4pt, topsep=-1pt, parsep=0pt]
\item \textbf{Heterogeneous Schema Extraction:} The agent first automatically detects and processes all available data sources, including SQLite databases, CSV files, JSON documents, and unstructured TXT files. It performs lightweight sampling on each source to extract metadata. For structured data (SQL, CSV), it extracts schemas, column names, data types, and representative example values. For unstructured text, it goes a step further by performing \textbf{\textit{content summarization and background knowledge expansion}} using an LLM to generate a ``pseudo-schema'' (e.g., a summary column), making the text's content discoverable and linkable.

\item \textbf{Unified Metadata Hub (MetaGraph) Construction:} All extracted metadata is centralized into a unified \textbf{MetaGraph}. In this graph, each column from every data source is treated as a node and assigned a globally unique alias (e.g., `csv.sales.user\_id', `sqlite.users.customer\_id'). Each node stores its properties, such as data type and example values. This creates a comprehensive, queryable catalog of the entire data ecosystem, providing a holistic data preview before any intensive processing occurs.

\textbf{Formalization of MetaGraph Construction:}
Let the set of all data sources be $\mathcal{D} = \{d_1, d_2, \dots, d_N\}$. For each source $d_i \in \mathcal{D}$, we extract a set of columns $\mathcal{C}_i = \{c_{i,1}, c_{i,2}, \dots, c_{i,m_i}\}$. Each column $c$ is formally represented as a tuple containing its unique identifier (alias), data type, and a set of sample values:
\begin{center}
    $c = (\text{id}(c), \tau(c), \mathcal{E}(c))$
\end{center}
where $\text{id}(c)$ is the unique alias, $\tau(c)$ is its data type (e.g., INT, TEXT, FLOAT), and $\mathcal{E}(c)$ is a set of sampled example values. The MetaGraph, $\mathcal{G}_M$, is then defined as the total set of all columns from all data sources:
\begin{center}
    $\mathcal{G}_M = \bigcup_{i=1}^{N} \mathcal{C}_i$
\end{center}

\item \textbf{Entity-Graph Generation via Similarity Analysis:} With the MetaGraph in place, the agent proceeds to discover potential relationships between columns. It performs a pairwise comparison of all columns in the MetaGraph, applying a similarity heuristic to identify potential join keys.

\textbf{Formalization of Similarity Analysis:}
We formalize the EntityGraph as a weighted, undirected graph $\mathcal{G}_E = (V, E, W)$, where the set of vertices $V$ is the MetaGraph itself ($V = \mathcal{G}_M$). The set of edges $E$ represents probable entity linkages, with an edge ${c_a, c_b}$ existing between any two distinct, type-compatible columns ($c_a, c_b \in V$ with $\tau(c_a) = \tau(c_b)$) if their similarity score $\text{Sim}(c_a, c_b)$ exceeds a predefined threshold $\theta$. The weight function $W$ assigns this similarity score to each edge, such that $W(c_a, c_b) = \text{Sim}(c_a, c_b)$. The similarity score is calculated as a weighted linear combination of name and value distribution similarity:
\begin{center}
$\text{Sim}(c_a, c_b) = w_n \cdot \text{Sim}_{\text{name}}(c_a, c_b) + w_v \cdot \text{Sim}_{\text{val}}(c_a, c_b)$
\end{center}
Here, $w_n$ and $w_v$ are hyperparameters representing the weights for name and value similarity, respectively (e.g., $w_n=0.6, w_v=0.4$), with $w_n + w_v = 1$. The name similarity component, $\text{Sim}{\text{name}}$, measures the lexical similarity between column names, using methods ranging from simple exact match indicators to more advanced metrics like Levenshtein or Jaro-Winkler distance. The value similarity, $\text{Sim}{\text{val}}$, is computed using the Jaccard similarity coefficient on the sets of sampled example values, $\mathcal{E}(c_a)$ and $\mathcal{E}(c_b)$:
\begin{center}
$\text{Sim}_{\text{val}}(c_a, c_b) = \frac{|\mathcal{E}(c_a) \cap \mathcal{E}(c_b)|}{|\mathcal{E}(c_a) \cup \mathcal{E}(c_b)|}$
\end{center}
The resulting EntityGraph, $\mathcal{G}_E$, thus maps potential join paths across the different data sources, with the edge weights quantifying the confidence of each linkage.

\item \textbf{Actionable Join-Hint Formulation:} Finally, the agent takes the highest-scoring relationships from the EntityGraph and formulates a concise, human-readable \textbf{Join-Hint}. This hint, such as \textit{"You must JOIN ON: csv.sales.user\_id = sqlite.users.customer\_id"}, serves as a direct instruction for the subsequent code generation phase.

\textbf{Formalization of Hint Selection:}
Let the set of discovered edges in the EntityGraph be $E$. The agent constructs an ordered list of candidate join keys, $\mathcal{L}$, by sorting the edges in $E$ in descending order of their weight $W$.
\begin{center}
    $\mathcal{L} = \text{sorted}(E, \text{key}=W, \text{reverse=True})$
\end{center}
The final set of hints, $\mathcal{H}$, is generated by selecting the top-$k$ edges from this list:
\begin{center}
    $\mathcal{H} = \{e \in \mathcal{L} \mid \text{rank}(e) \le k\}$
\end{center}
This set $\mathcal{H}$ is then translated into the textual Join-Hint provided to the code generation module.

\end{enumerate}

\subsubsection{Hierarchical Planning with ReAct-style Reasoning}
At its core, ReActInsight leverages a hierarchical planning module that operationalizes the ReAct paradigm. Given a high-level analytical objective from the user (e.g., \textit{``Analyze the recent drop in sales''}), the agent first reasons about the potential causes and breaks the problem down into a sequence of logical sub-goals. As illustrated in Figure~\ref{fig:Architecture}, this hierarchical decomposition transforms a broad query into a structured analytical plan. Each sub-goal is then further decomposed into specific, actionable sub-questions (e.g., \textit{``What are the total sales per product category for the last quarter?''}), which directly guide the code generation step.

\begin{table*}[!t]
\centering
\resizebox{0.95\linewidth}{!}{
\begin{tabular}{lcccccccc}
\toprule
\multirow{2}{*}{\bf Agent} & \multicolumn{4}{c}{\bf Insight\,-\,level Scores (G-Eval)} & \multicolumn{4}{c}{\bf Summary\,-\,level Scores (G-Eval)} \\[-0.5mm]
\cmidrule(r){2-5}\cmidrule(r){6-9}
& \bf Easy & \bf Medium & \bf Hard & \bf Avg. & \bf Easy & \bf Medium & \bf Hard & \bf Avg. \\[0.5mm]
\midrule
ReAct \cite{yao2023react} & 0.3115 & 0.3024 & 0.2982 & 0.3040& 0.3549 & 0.3430 & 0.3223 & 0.3401 \\
CodeGen \cite{majumder2024discoverybench} & 0.3197 & 0.2999 & 0.3256 & 0.3151& 0.3796 & 0.3342 & 0.4075 & 0.3738 \\
Pandas Agent \cite{langchain2024pandas} & 0.3831 & 0.3978 & 0.3740 & 0.3850 & 0.5551 & 0.5332 & 0.4905 & 0.5263 \\
D2D \cite{zhang2025data}  & 0.3890 & 0.4346 & 0.3234 & 0.3823 & 0.4964 & 0.5235 & 0.4336 & 0.4845 \\
DABstep \cite{egg2025dabstepdataagentbenchmark} & 0.4236 &  0.4326 &  0.3916 & 0.4159 &  0.5379 & 0.5377 &  0.4744 & 0.5167 \\
AgentPoirot \cite{sahu2024insightbench} & 0.4810 & 0.4448 & 0.3826 & 0.4361 & 0.5917 & 0.5708 & 0.5229 & 0.5618 \\
\rowcolor{bluei}  ReActInsight (Ours) & \textbf{0.5030} & \textbf{0.4704} & \textbf{0.4517} & \textbf{0.4750} & \textbf{0.5984} & \textbf{0.5789} & \textbf{0.5605} & \textbf{0.5792} \\

\rowcolor{bluei} \hspace{10px} \textit{Improvement (\%)} & +4.57\% & +5.76\% & +18.06\% & +8.92\% & +1.13\% & +1.42\% & +7.19\% & +3.10\% \\
\bottomrule
\end{tabular}
}

\caption{Comparison of agent performance across varying difficulty levels on {\sc UniDataBench}.}
\vspace{-5mm}
\label{tab:agent_insight_scores_mainbody}
\end{table*}

\subsubsection{Robust Code Generation and Adaptive Visualization}
For each actionable sub-question, ReActInsight generates the necessary code (e.g., SQL queries, Python scripts using Pandas) to perform the analysis. Crucially, the prompt provided to the LLM for code generation is augmented with the \textbf{Join-Hint} derived from the EntityGraph. This explicitly guides the model to construct correct JOIN statements (e.g., SQL JOINs or Python Pandas merges) across different data sources, which is essential for answering complex, multi-faceted questions.

This process is designed to be robust. After generation, the agent executes the code in a sandboxed environment and captures all outputs, including stdout and stderr. If an error is detected, a self-correction loop is initiated. The agent feeds the original code, the error message, and the corresponding sub-goal back into itself to generate a corrected version, iterating until the code executes successfully. Upon successful execution, a key innovation of our agent becomes apparent: its adaptive visualization capability. Instead of defaulting to simple histograms or tables, the agent reasons about \textbf{\textit{the nature of the data (now potentially joined from multiple sources) and the sub-question}} to select the most appropriate visualization type to reveal underlying patterns. This includes generating scatter plots to show correlations, time-series plots for trends, heatmaps for matrix data, and bar charts for comparisons, ensuring that the intermediate results are not only computationally correct but also intuitively presented.

\subsection{Insight Synthesis and Summary}
ReActInsight does not simply present a series of charts; it analyzes the results from each successfully executed step. It aggregates the observations, identifies key findings, and constructs a final, coherent narrative. This summary report provides users with a high-level overview, detailed findings supported by visualizations, and actionable recommendations, fulfilling end-to-end data analysis.

\subsection{Optimization via Model Cascading}
To balance performance with operational costs, ReActInsight implements a model cascading strategy. Routine and structurally simple tasks, such as the generation of sub-questions, are delegated to smaller, faster, and more cost-effective models (e.g., GPT-3.5-Turbo). Only when a task proves too complex,\textbf{\textit{ or when the self-correction module is triggered by repeated failures,}} does the agent escalate the task to a more powerful state-of-the-art model (e.g., GPT-4o, GPT-5). This strategy significantly optimizes both computational cost and response latency, making ReActInsight a practical and efficient solution for real-world applications.

\section{Experiments}

\begin{figure*}[ht!]
    \centering
    \begin{minipage}[b]{0.62\linewidth}
        \includegraphics[width=\linewidth]{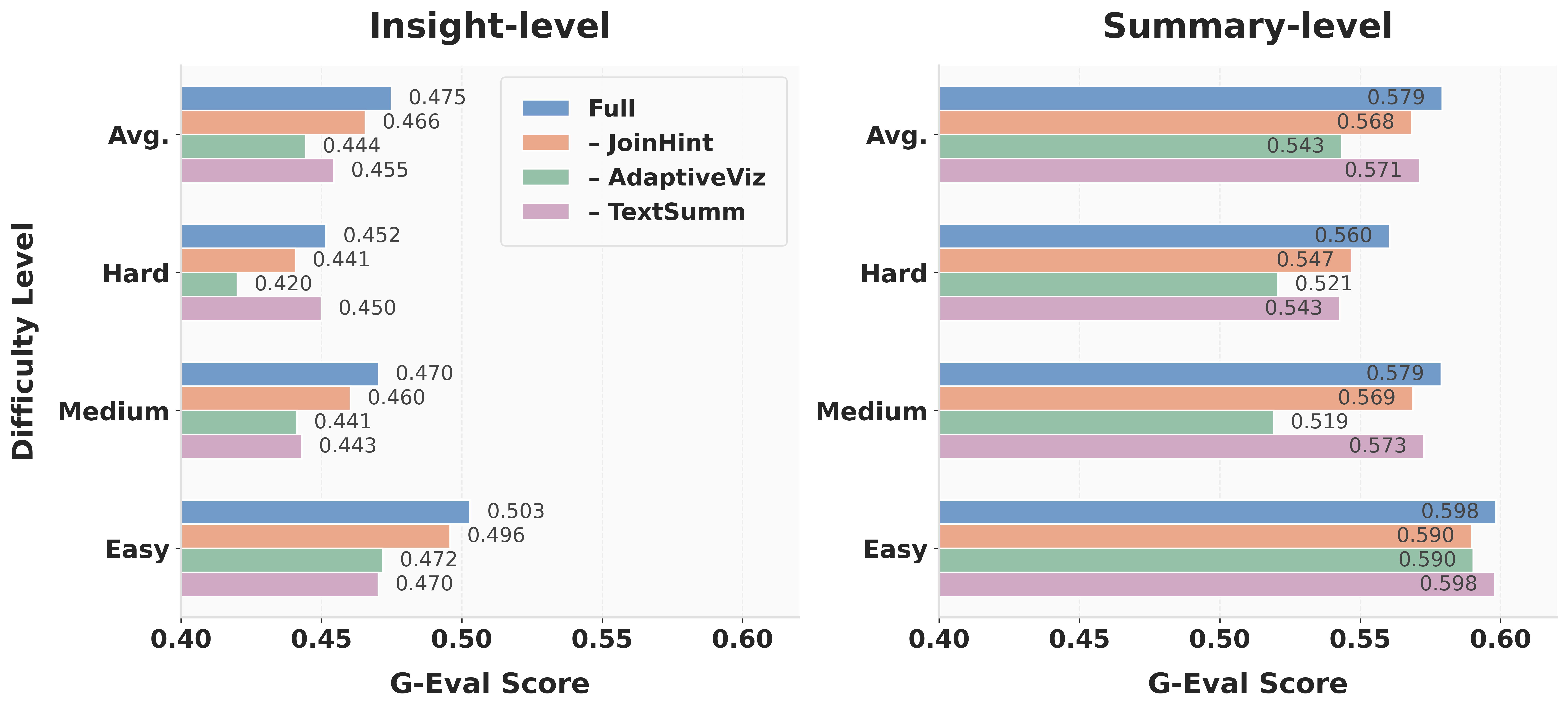}
        \vspace{-15pt}
        \caption{Ablation Study Results.}
        \label{fig:ablation_results}
    \end{minipage}\hfill
        \begin{minipage}[b]{0.38\linewidth}
        \includegraphics[width=\linewidth]{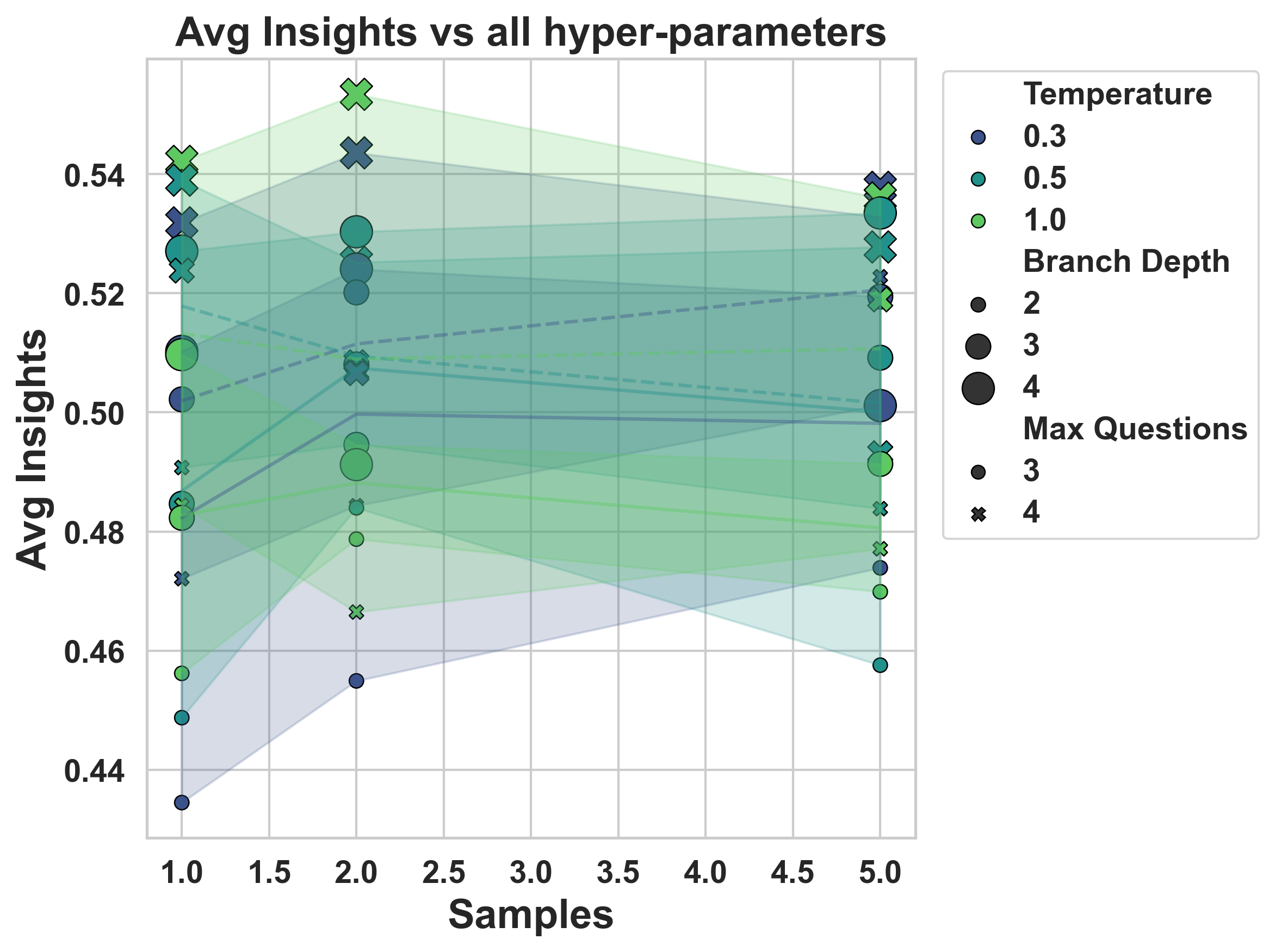}
        \vspace{-15pt}
        \caption{Hyperparameter Study.}
        \label{fig:hyper}
    \end{minipage}\hfill
    \vspace{-13pt}
\end{figure*}

\subsection{Experimental Setup}

\noindent \textbf{Baselines.}
We benchmarked a diverse set of state-of-the-art data analytics agents on UniDataBench, including CodeGen~\cite{majumder2024discoverybench}, ReAct~\cite{yao2023react}, Data-to-Dashboard~\cite{zhang2025data}, Pandas Agent~\cite{langchain2024pandas}, DABstep~\cite{egg2025dabstepdataagentbenchmark}, and AgentPoirot~\cite{sahu2024insightbench}. We adapted them to handle our multiple data sources for a fair comparison. Detailed descriptions of all baselines are provided in Appendix~\ref{appendix:baselines}.

\noindent \textbf{Evaluation Metrics.}
Following established practices~\cite{sahu2024insightbench}, we use the G-Eval~\cite{liu-etal-2023-g} metric to assess agents at both level of insight and summary. More detailed definition of metrics are provided in Appendix \ref{appendix:metric}.

\noindent \textbf{Implementation Details.}
We involved a comparative analysis across a range of LLMs to serve as the backbone for our agents. For all experiments, a temperature of 0 was applied. The prompts employed are provided in Appendix \ref{appendix:prompts}.

\subsection{Performance Comparison}

As detailed in Table~\ref{tab:agent_insight_scores_mainbody}, ReActInsight surpasses all baseline methods across all difficulty levels on the { UniDataBench} benchmark. Single-agent methods, CodeGen and ReAct demonstrate limited effectiveness, highlighting the inadequacy of generic code generation or simple action-based frameworks for complex, multi-source data analysis. The specialized focus of Data-to-Dashboard and Pandas Agent on data analysis leads to respectable scores and AgentPoirot's multi-step sub-question approach provides further improvement. However, their performance difference is primarily on ``Easy'' difficulty tasks, with smaller variations on more challenging tasks, underscoring the need for optimized multi-source data processing. Furthermore, the mediocre results of DABstep, an agent designed for multi-source analysis, indicate that existing multi-source agents are not easily adapted to high-dimensional data analysis. %
Our proposed method, ReActInsight, demonstrates superior performance by initially constructing a unified map of all data sources, which is then combined with hierarchical planning and self-correction. ReActInsight achieves an insight-level score of 0.4750 and a summary-level score of 0.5792, representing an 8.9\% and 3.1\% improvement over the baseline, respectively.
Notably, the performance advantage of ReActInsight becomes more pronounced as the task difficulty increases. On ``hard'' difficulty tasks, our model scores 0.4517, surpassing the AgentPoirot baseline (0.3826) by a significant 18\%. This trend highlights our agent capability in handling data analysis tasks that involve a variety of data formats.
Case studies are presented in Appendix~\ref{appendix:case_study}.

\subsection{Ablation Study}
We conducted an ablation study by systematically removing the JoinHint mechanism, the Adaptive Visualization module, and the Text Summarizer from the full ReActInsight (gpt-4o). As shown in Figure~\ref{fig:ablation_results}, the results unequivocally demonstrate that the full framework outperforms all ablated variants across both Insight-level and Summary-level evaluations, confirming that each component contributes positively to the overall performance.

The removal of the Adaptive Visualization module caused the most significant performance degradation, with the average Summary-level score dropping from 0.5792 to 0.5434. This highlights its critical role in generating effective charts for deriving insights. Ablating the Text Summarizer also led to a notable decline, while removing the JoinHint mechanism resulted in a more modest but still consistent performance drop.

\subsection{Hyperparameter Study} %

We conducted a hyperparameter study to evaluate the impact of \texttt{temperature, samples, max questions, and branch depth} on the quality of generated insights and summaries. As shown in Figure \ref{fig:hyper}, a fundamental trade-off between the two objectives, demonstrating that no single configuration is optimal for both. Specifically, the highest insight score (0.5534) was achieved with a configuration favoring creativity and deep exploration: a high \texttt{temperature} (1.0), \texttt{max questions} of 4, and a \texttt{branch depth} of 4. Conversely, in Figure \ref{fig:hyper_sum}, the peak summary score (0.6001) required a setting that prioritizes precision and focus: a low temperature (0.3), a high sample count (\texttt{samples}=5), and a more moderate exploration depth (\texttt{max questions}=3, \texttt{branch depth}=3). This clear divergence in optimal settings validates the approach of employing task-specific hyperparameter configurations to maximize performance for each respective goal. 

\section{Conclusion}
In this paper, we presented {UniDataBench}, a new benchmark for evaluating data analytics agents' ability to work with structured and unstructured data. {UniDataBench} offers a diverse set of datasets and tasks reflecting real-world business analytics challenges. We also proposed {ReActInsight}, an LLM-based agent designed to handle multiple data sources. 
Our experiments show {UniDataBench} effectively assesses agent performance, highlighting its potential to drive innovation in data analytics. 
Overall, our work provides a robust foundation for advancing data analytics agents in handling complex, real-world data environments.

\section*{Limitations}
The current data construction approach relies heavily on manual construction, which is labor-intensive and costly. Moreover, the scale of the resulting datasets is limited due to the constraints of human effort. Therefore, the next step will explore automated construction methods to overcome these limitations and expand the dataset size efficiently.
At present, the dataset supports unstructured and structured data types. It lacks support for multimodal data, which includes a combination of text, images, audio, and other forms of data. To address this gap, the next phase will focus on developing a multimodal dataset to better reflect real-world data complexity and enhance the applicability of the dataset. The existing dataset is limited in its coverage of diverse domains and scenarios. It primarily focuses on a limited range of applications, which may not be representative of broader use cases. To improve the generalizability and utility of the dataset, future work will aim to expand its domain coverage and include a wider variety of scenarios to better support various research and development needs.

\section*{Ethical Statement}
Our agent is primarily designed to assist users in automatically generating data analysis reports and uncovering valuable insights from diverse data sources. However, like all AI systems, it is not infallible. For instance, LLMs and agents may sometimes generate hallucinations or produce misleading information. Therefore, the reports generated by our system should only be used as an auxiliary tool. We strongly recommend that professionals thoroughly review and verify the results before making any critical decisions.

\bibliography{custom}

\onecolumn
\newpage
\twocolumn
\appendix

\begin{figure*}[t!]
    \centering
    \includegraphics[width = 1\textwidth]{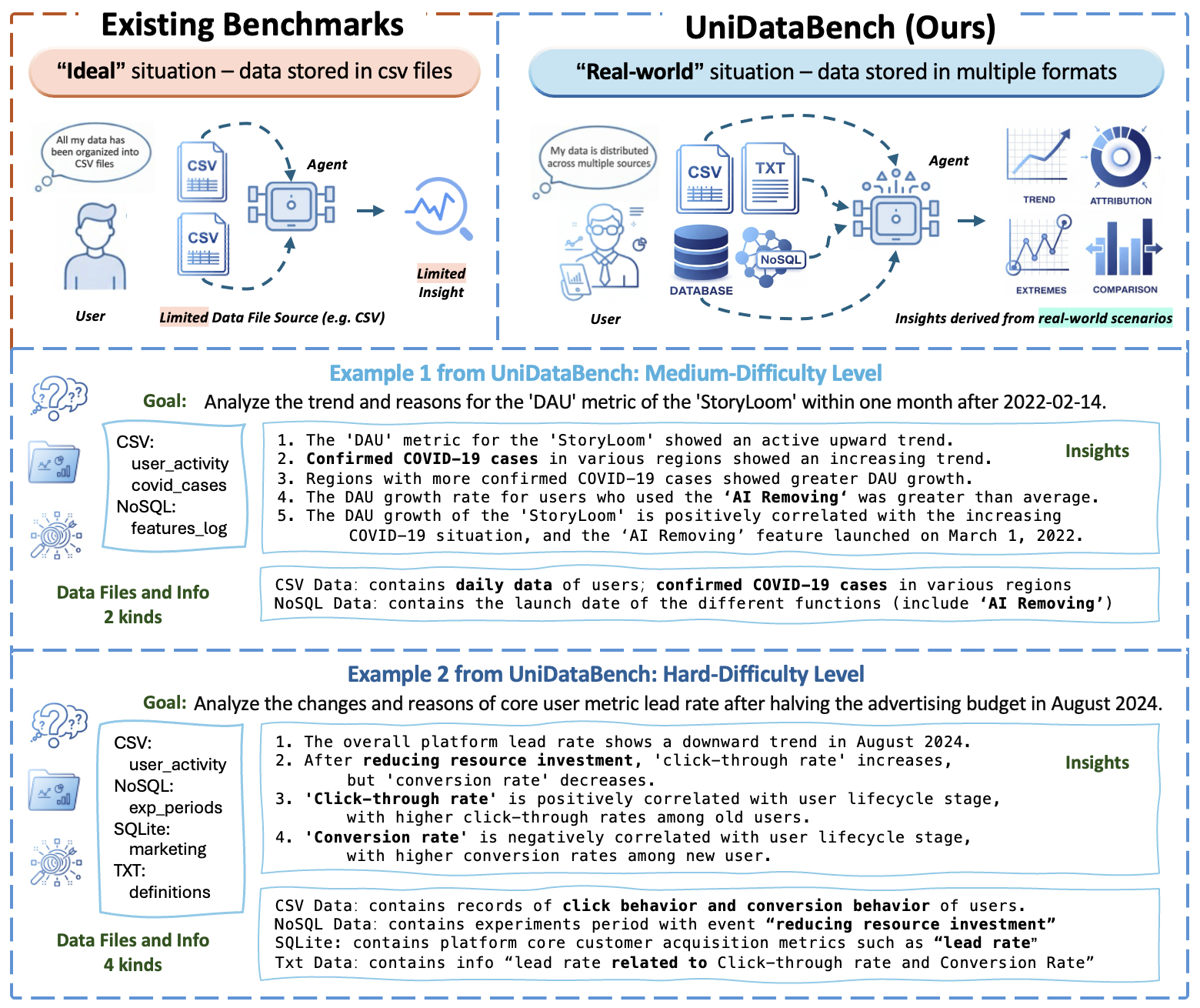}
    \vspace{-20pt}
    \caption{Existing benchmarks (left) that evaluate agents on isolated data sources. This represents an idealized yet limited scenario for data analysis, which leads to incomplete insights. UniDataBench (right), designed to simulate practical data analysis requirements. User pose a high-level goal, then will get comprehensive insights about trends, extreme values, comparisons and attribution.
    }
    \label{fig:caption_figure}
\end{figure*}

\section{UniDataBench Details}

\subsection{Task Illustration}
As illustrated in the top of Figure \ref{fig:caption_figure}, existing benchmarks often rely on idealized scenarios with data confined to a single format, such as CSV files. In contrast, UniDataBench provides a "real-world" situation where data is stored in multiple formats (e.g., CSV, TXT, NoSQL, SQLite), compelling the agent to integrate information to fulfill a user's analytical goal.
Each problem instance in UniDataBench is structured to test this capability and consists of three key components:
\begin{itemize}
[leftmargin=*,itemsep=2pt,topsep=0pt,parsep=0pt]
\item \textbf{Goal}: A high-level, open-ended question or objective posed by the user.
\item \textbf{Data Files and Info}: A collection of data files in multiple sources.
\item \textbf{Ground-Truth Insights}: A set of key insights that an agent is expected to discover.
\end{itemize}

\subsection{Typical Examples in UniDataBench}

To provide a concrete understanding of our benchmark's design, we illustrate its structure through two examples, as shown in Figure \ref{fig:caption_figure}.
\paragraph{Example 1: Medium-Difficulty Level}
This task simulates a typical business intelligence query requiring trend analysis and attribution.
\begin{itemize}
[leftmargin=*,itemsep=2pt,topsep=0pt,parsep=0pt]
\item \textbf{Goal}: Analyze the trend and reasons for the 'DAU' (Daily Active Users) of the 'StoryLoom' feature within one month after 2022-02-14.
\item \textbf{Data Files}: The task involves three data files across two formats:
\begin{itemize}
\item CSV: \texttt{user\_activity}, \texttt{covid\_cases}
\item NoSQL: \texttt{features\_log}
\end{itemize}
\item \textbf{Analytical Challenge}: To successfully answer the query, an agent cannot rely on a single file. It must first identify the DAU trend from \texttt{user\_{activity}} data. Then, to explain the reasons for this trend, it must integrate external factors from \texttt{covid\_cases} and internal product changes from the \texttt{features\_log} (which contains the launch date of a new 'AI Removing' function). The expected insights involve correlating the DAU growth with both the rising COVID-19 cases and the launch of the new feature.
\end{itemize}

\paragraph{Example 2: Hard-Difficulty Level}
This task presents a more complex scenario involving a deeper causal analysis across a wider array of data types, including unstructured text and a relational database.
\begin{itemize}
[leftmargin=*,itemsep=2pt,topsep=0pt,parsep=0pt]
\item \textbf{Goal}: Analyze the changes and reasons of the core user metric 'lead rate' after halving the advertising budget in August 2024.
\item \textbf{Data Files}: This problem includes data from four different formats:
\begin{itemize}
\item CSV: \texttt{user\_activity}
\item NoSQL: \texttt{exp\_periods}
\item SQLite: \texttt{marketing}
\item TXT: \texttt{definitions} 
\end{itemize}
\item \textbf{Analytical Challenge}: The complexity is significantly increased. The agent must first consult the unstructured \texttt{definitions.txt} file to understand the key metric. It then needs to query the SQLite database (\texttt{marketing}) and join this information with user behavior data from CSV (\texttt{user\_activity}) and event timelines from NoSQL (\texttt{exp\_periods}). The core challenge lies in disentangling the effects of the budget cut on different metrics (click-through rate vs. conversion rate) and segmenting the analysis by user lifecycle to uncover nuanced insights. This requires the agent to not only handle more data formats but also to construct a more sophisticated analytical workflow.
\end{itemize}
Through these carefully constructed tasks, UniDataBench provides a robust framework for evaluating an agent's ability to perform integrated, multi-faceted data analysis that is representative of practical, real-world requirements.

\subsection{Example of Synthesizing and Validating a Multi-Source Dataset}
\label{sec:appendix_synth_example}

\noindent We outline a practical case of how we created a multi-source dataset for a complex analytical scenario. This example start from following goal:

\vspace{0.5em}
\noindent\textit{\textbf{Goal:} Analyze the combined impact of the COVID-19 pandemic and advertising budget cuts on the lead conversion rate (CVR) in Q2 2022.}
\vspace{0.5em}

\noindent\textit{\textbf{Insights we aim to embed within the data}}:
\begin{itemize}[leftmargin=*,itemsep=2pt,topsep=0pt,parsep=0pt]
    \item The lead conversion rate (CVR) in Q2 2022 showed an overall downward trend.
    \item The CVR is negatively correlated with the severity of the pandemic.
    \item After implementing an advertising budget cut, the rate of CVR decline slowed down.
    \item The improvement in user quality from reduced ad investment mitigated the overall CVR decline.
\end{itemize}

\subsubsection{Schema Design and Standardization}

\noindent Based on the goal and ground-truth insights, we first design the schemas for the required multi-source data. This involves defining core fields that directly enable the analysis and ensuring relationships are maintained across files.

\vspace{0.5em}

\noindent \textit{\textbf{(a) Core Analysis Dataset (CSV):}} This file contains the primary time-series data. The schema is designed to directly support the core analysis by linking CVR to the pandemic and ad spend.

\begin{table}[h!]
    \centering
    \small
    \begin{tabular}{@{}lll@{}}
        \toprule
        \textbf{Field Name} & \textbf{Type} & \textbf{Description} \\
        \midrule
        log\_date & string & Log date for Q2 2022 \\
        region\_id & string & Unique region identifier\\
        daily\_cvr & float & Daily conversion rate \\
        ad\_spend & float & Daily advertising spend in USD \\
        \dots & \dots & \textit{(12 other columns)}\\
        \bottomrule
    \end{tabular}
    \caption{Schema for the core `daily\_summary.csv` file.}
    \label{tab:daily_summary_schema}
\end{table}

\noindent \textit{\textbf{(b) Event Period Data (NoSQL):}} The time boundaries for the advertising strategy change are stored in NoSQL. This simulates a real-world scenario where company-wide events are logged in a semi-structured format. The agent must parse this file to find the relevant periods—"Investment Period" and "Budget Cut Period"—among other irrelevant entries, and extract their start and end dates to correctly segment the analysis.

\begin{lstlisting}[language=json, style=jsonstyle, caption={Excerpt from `company\_events.json`. The agent must identify the relevant financial periods among other distracting entries.}, label=json_example]
{
  "company_periods": [
    {
      "period_id": "p2021q4-sales",
      "period_name": "Q4 2021 Holiday Sale",
      "type": "Marketing",
      ...
    },
    {
      "period_id": "p2022q1-invest",
      "period_name": "Investment Period",
      "type": "Financial",
      "start_date": "2022-03-01",
      "end_date": "2022-05-31",
      "description": "Period of increased spending..."
    },
    {
      "period_id": "p2022q2-budget",
      "period_name": "Budget Cut Period",
      "type": "Financial",
      "start_date": "2022-06-01",
      "end_date": "2022-06-30",
      "description": "A short period of reduced expenditure..."
    }
  ]
}
\end{lstlisting}

\vspace{1em}

\subsubsection{Distractor Column Introduction}

\noindent To enhance analytical complexity, we introduce more than 10 distractor columns into the `csv` file (e.g., \texttt{campaign\_id}, \texttt{weather\_index}). Similarly, there are some  irrelevant event periods (e.g., "Q4 2021 Holiday Sale") stored in `nosql` file. This multi-source noise challenges the agent to perform effective feature selection and information extraction, a common task for data analysts.

\subsubsection{Synthetic Data Creation}

\noindent The entire data generation pipeline is implemented using Python scripts. We instruct an LLM to generate a script that adheres to the designed schemas and embeds the intended patterns. For example, the script defines region-specific parameters to control the base CVR and its sensitivity to factors like COVID-19 severity and ad spend.

\begin{lstlisting}[language=Python, frame=single, caption={Python snippet for generating region-specific metadata used in data synthesis.}, label={lst:python_code_example}, basicstyle=\ttfamily\small]
# Create region metadata
region_meta = []
for i in range(NUM_REGIONS):
    region_id = f"R{1000 + i}"
    region_name = f"Region-{chr(65 + (i %
    # base cvr for region (between 40%
    base_cvr = random.uniform(0.4, 0.60)
    # base daily ad spend (mean) in dollars
    base_ad_spend = random.uniform(800, 4500)
    # region-level sensitivity to covid severity (coeff)
    covid_sensitivity = random.uniform(0.0009, 0.0022)
    # ad spend effect on CVR (higher spend, lower quality)
    ad_spend_effect = random.uniform(0.00002, 0.00005)
    region_meta.append({
        "region_id": region_id,
        "region_name": region_name,
        "base_cvr": base_cvr,
        "base_ad_spend": base_ad_spend,
        "covid_sensitivity": covid_sensitivity,
        "ad_spend_effect": ad_spend_effect
    })
\end{lstlisting}

\subsubsection{Data Validation and Visualization}

\begin{figure*}[t!]
    \centering
    \begin{subfigure}[b]{0.64\textwidth}
\includegraphics[width=\linewidth]{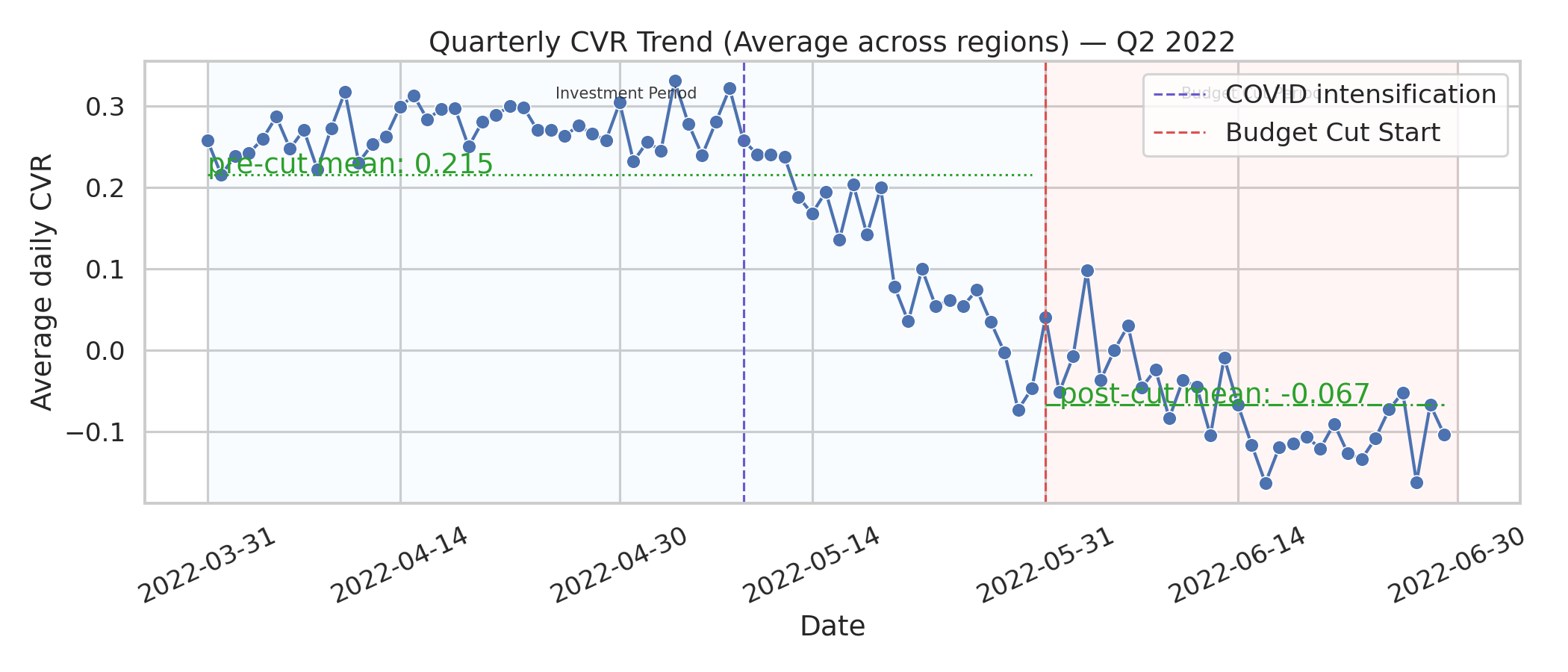}
\caption{Time series analysis of conversion rate.}
\label{fig:cvr_timeseries}
    \end{subfigure}
    \hfill
    \begin{subfigure}[b]{0.34\textwidth}
\includegraphics[width=\linewidth]{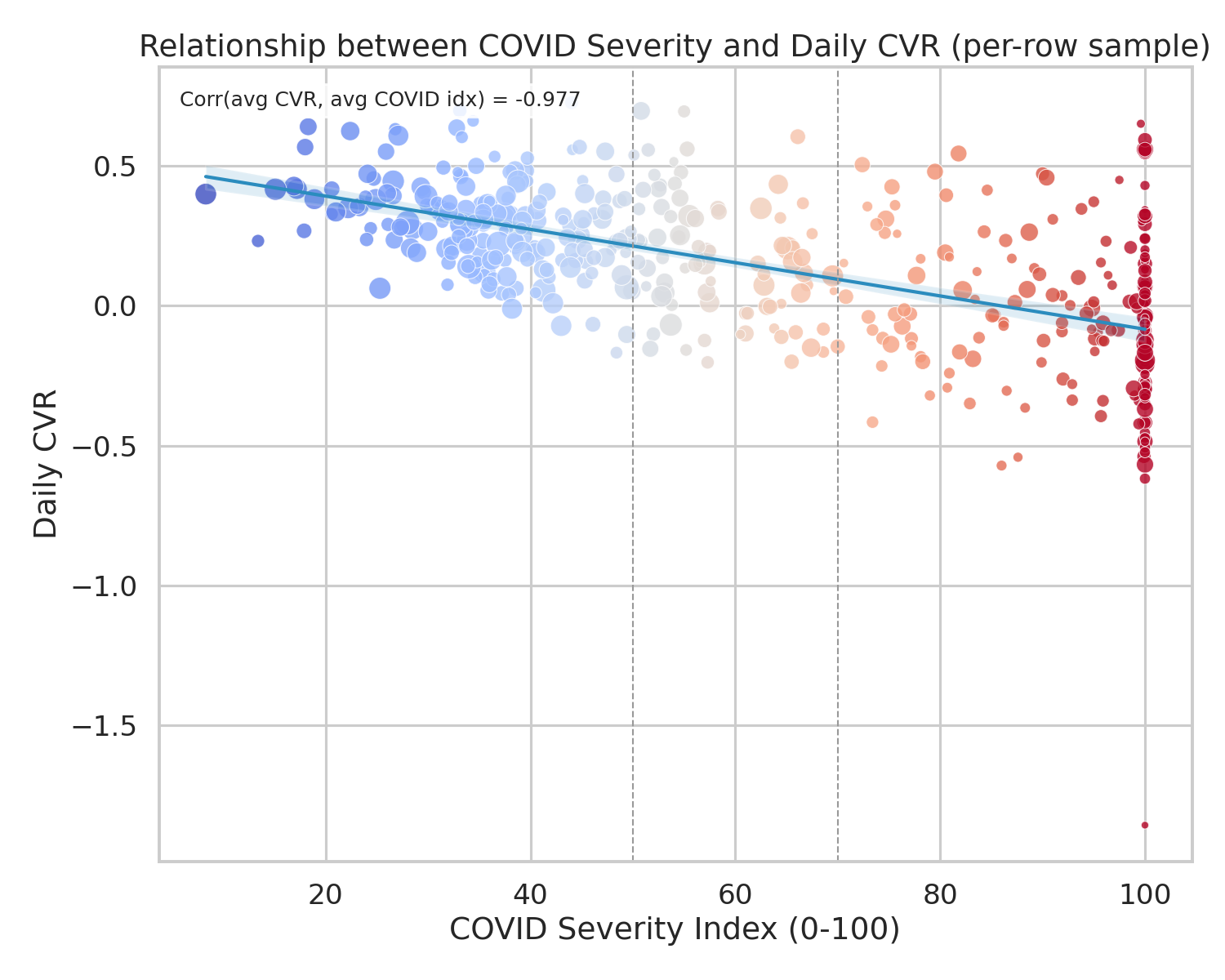}
\caption{Rate versus COVID severity.}
\label{fig:cvr_vs_covid}
    \end{subfigure}
\caption{Validation plot for the synthesized data, confirming that the intended CVR trends and the impact of the budget cut are successfully embedded.}
    \label{fig:validation_plots}
\end{figure*}

After generation, we perform a crucial validation step to ensure the ground-truth insights are discoverable. Visualization is a primary method for this confirmation. As shown in Figure~\ref{fig:validation_plots}, the plot serves as visual proof that the intended analytical patterns are present and accessible. Specifically, the visualization confirms: (1) the overall downward trend of CVR, (2) its negative correlation with the pandemic's severity, and (3) a clear change in the CVR's decline rate after the budget cut was implemented in June.

\subsection{Task Across Analysis Theme}
To ensure the practical relevance and comprehensive coverage of our benchmark, the tasks in UniDataBench are sourced from authentic, real-world data analysis scenarios. These scenarios are thoughtfully categorized into five core business themes, each representing a common area of analytical inquiry in industry.
The analytical tasks are structured across five core business domains: User Behavior Analysis (29 tasks), Business Performance Analysis (23 tasks), Market Competition Analysis (13 tasks), Sales Channel Analysis (18 tasks), and Macro-Industry Analysis (17 tasks). An illustrative example for each of these five themes is provided in Table~\ref{tab:domain_category_example}, showcasing a representative goal, the expected insights, and the necessary data files for each task.

\label{sec:appendix_data_statistics}
\begin{table*}[h!]
\centering
\scriptsize
\renewcommand{\arraystretch}{1}
\begin{tabular}{p{0.2\linewidth}p{0.6\linewidth}p{0.2\linewidth}}
\toprule
\textbf{Goal} & \textbf{Insights} & \textbf{Data Files}  \\
\midrule
\multicolumn{3}{l}{\textbf{Task Domain:} User Behavior Analysis (Tasks Size: 29)}\\
\multicolumn{3}{l}{\textbf{Task No.:} 1, 2, 3, 4, 5, 6, 8, 9, 10, 11, 12, 13, 14, 15, 16, 21, 22, 23, 24, 25, 42, 48, 49, 53, 54, 55, 56, 85, 92}\\
\multicolumn{3}{l}{%
  \tikz[remember picture, baseline=-0.5ex] \draw[dashed, line width=0.4pt] 
    ($(0,0)$) -- ($(1.1\linewidth,0)$);
} \\
\parbox[t]{\linewidth}{\ttfamily 
Analyze the impact of features launched in Q2 2024 on the 'DAU' metric for the video editing product 'StoryLoom'
}
& \parbox[t]{\linewidth}{\ttfamily 
1. The overall DAU metric showed a rising trend in Q2 2024.\\
2. The DAU of users who used the 'Fill Light Filter' feature accounts for a high proportion of the overall DAU, while the DAU of users who used the 'Vintage Filter' feature accounts for a very low proportion.\\
3. The DAU for users of the 'Fill Light Filter' feature showed a rising trend.\\
4. The DAU for users of the 'Vintage Filter' feature showed no significant change.\\
5. The increase in DAU this quarter is mainly attributed to the positive impact of the 'Fill Light Filter' feature, whereas the 'Vintage Filter' feature had no significant impact on the overall DAU growth.}
& \parbox[t]{\linewidth}{\ttfamily features\_metadata.json \\
storyloom\_usage\_q2.csv}\\
\midrule
\multicolumn{3}{l}{\textbf{Task Domain:} Business Performance Analysis (Tasks Size: 23)}\\
\multicolumn{3}{l}{\textbf{Task No.:} 7, 26, 27, 28, 32, 33, 34, 37, 38, 39, 40, 43, 44, 45, 46, 47, 51, 73, 75, 91, 93, 94, 99}\\ 
\multicolumn{3}{l}{%
  \tikz[remember picture, baseline=-0.5ex] \draw[dashed, line width=0.4pt] 
    ($(0,0)$) -- ($(1.1\linewidth,0)$);
} \\
\parbox[t]{\linewidth}{\ttfamily Analyze the changes in rental business transaction volume and the reasons from February to July 2023}
& \parbox[t]{\linewidth}{\ttfamily 1. The transaction volume of the rental business exhibited an upward trend from February to July.\\
2. The total exposure of the rental business has maintained a continuous upward trend since March 2023.\\
3. The transaction volume of all housing types showed an upward trend.\\
4. The new housing type 'Branded Apartments' launched in April 2023 boosted the transaction volume of the rental business.\\
5. The upward trend of trading volume in June and July 2023 was significantly higher than before.
\\
6. The increase in the transaction volume of the rental business was jointly influenced by three factors: the rise in exposure since March, the launch of the new 'Branded Apartments' in April, and the peak rental season in June and July.}
& \parbox[t]{\linewidth}{\ttfamily Break\_Information\_Report.txt\\
daily\_transactions.csv\\
housing\_type\_daily.csv
} \\
\midrule
\multicolumn{3}{l}{\textbf{Task Domain:} Market Competition Analysis (Tasks Size: 13)}\\
\multicolumn{3}{l}{\textbf{Task No.:} 19, 20, 57, 58, 59, 60, 61, 62, 63, 64, 65, 66, 84} \\
\multicolumn{3}{l}{%
  \tikz[remember picture, baseline=-0.5ex] \draw[dashed, line width=0.4pt] 
    ($(0,0)$) -- ($(1.1\linewidth,0)$);
} \\
\parbox[t]{\linewidth}{\ttfamily Analyze Changes in Public Opinion on 'LucidPic' Membership Benefits in 2024 Based on NSR and Its Components}
& \parbox[t]{\linewidth}{\ttfamily
1. The NSR (Net Sentiment Ratio) of public opinion related to 'LucidPic' membership benefits improved from negative in Q3 to positive in Q4.\\
2. The number of positive feedback in Q3 was the lowest of the year, while the number of negative feedback was the highest of the year.\\
3. The number of positive feedback in Q4 was the highest of the year, while the number of negative feedback was the lowest of the year.\\
4. Among all quarters in 2024, only Q4 saw the number of positive feedback exceed that of negative feedback.\\
5. The increase in the number of positive feedback in Q4 was mainly concentrated in the aspect of 'Paid Features Valuable'.\\
6. The growth of positive feedback in Q4 was related to the paid feature 'AI removal' launched in Q4.}
& \parbox[t]{\linewidth}{\ttfamily lucidpic\_knowledge.txt\\
feature\_launch.json\\
membership\_feedback.db}
\\
\midrule
\multicolumn{3}{l}{\textbf{Task Domain:} Sales Channel Analysis (Tasks Size: 18)}\\
\multicolumn{3}{l}{\textbf{Task No.:} 17, 18, 29, 30, 31, 35, 36, 41, 50, 52, 83, 86, 87, 95, 96, 97, 98, 100} \\
\multicolumn{3}{l}{%
  \tikz[remember picture, baseline=-0.5ex] \draw[dashed, line width=0.4pt] 
    ($(0,0)$) -- ($(1.1\linewidth,0)$);
} \\
\parbox[t]{\linewidth}{\ttfamily Comparison of the conversion funnel efficiency between the 'QuickMart' platform APP and Mini-Program channels, identifying their respective strengths and bottlenecks
}
& \parbox[t]{\linewidth}{\ttfamily 1. The total order volume of the APP channel is higher than that of the 'Mini-Program' channel.\\
2. By comparing stages in the conversion funnel, the 'Mini-Program' has a higher 'Store\_Entry' stage conversion rate.\\
3. By comparing stages in the conversion funnel, the APP channel's 'Order\_Completion' stage conversion rate is significantly higher than that of the 'Mini-Program'.}
& \parbox[t]{\linewidth}{\ttfamily daily\_store\_performance.csv\\
funnel\_knowledge.txt
}\\
\midrule
\multicolumn{3}{l}{\textbf{Task Domain:} Macro-Industry Analysis (Tasks Size: 17)}\\
\multicolumn{3}{l}{\textbf{Task No.:} 67, 68, 69, 70, 71, 72, 74, 76, 77, 78, 79, 80, 81, 82, 88, 89, 90} \\
\multicolumn{3}{l}{%
  \tikz[remember picture, baseline=-0.5ex] \draw[dashed, line width=0.4pt] 
    ($(0,0)$) -- ($(1.1\linewidth,0)$);
} \\
\parbox[t]{\linewidth}{\ttfamily Analyze the indicator trends of PPI and CPI in developed economies from 2022 to 2024 and their impact on corporate profits}
& \parbox[t]{\linewidth}{\ttfamily 
1. From 2022 to 2024, both PPI and CPI showed a downward trend.\\
2. From 2022 to 2024, the decline rate of PPI was much greater than that of CPI.\\
3. Due to the difference in decline speeds, PPI remained significantly lower than CPI for most of the time, forming a persistent negative price spread.\\
4. The continuous widening of the negative PPI-CPI spread indicates that the cost of corporate inputs is falling faster than product selling prices, which is favorable for corporate profit margin expansion.
}
& \parbox[t]{\linewidth}{\ttfamily 
indicators\_definition.txt\\
monthly\_inflation\_data.csv}\\
\bottomrule
\end{tabular}
\caption{Examples for five distinct analysis categories from the Unidatabench benchmark. Each task includes a goal, serveral insights and some necessary data files.}
\vspace{-10pt}
\label{tab:domain_category_example}
\end{table*}

\section{Benchmark Quality Assessment}
\label{sec:quality_assessment}

\begin{figure*}[h!] 
    \centering \includegraphics[width=0.95\textwidth]{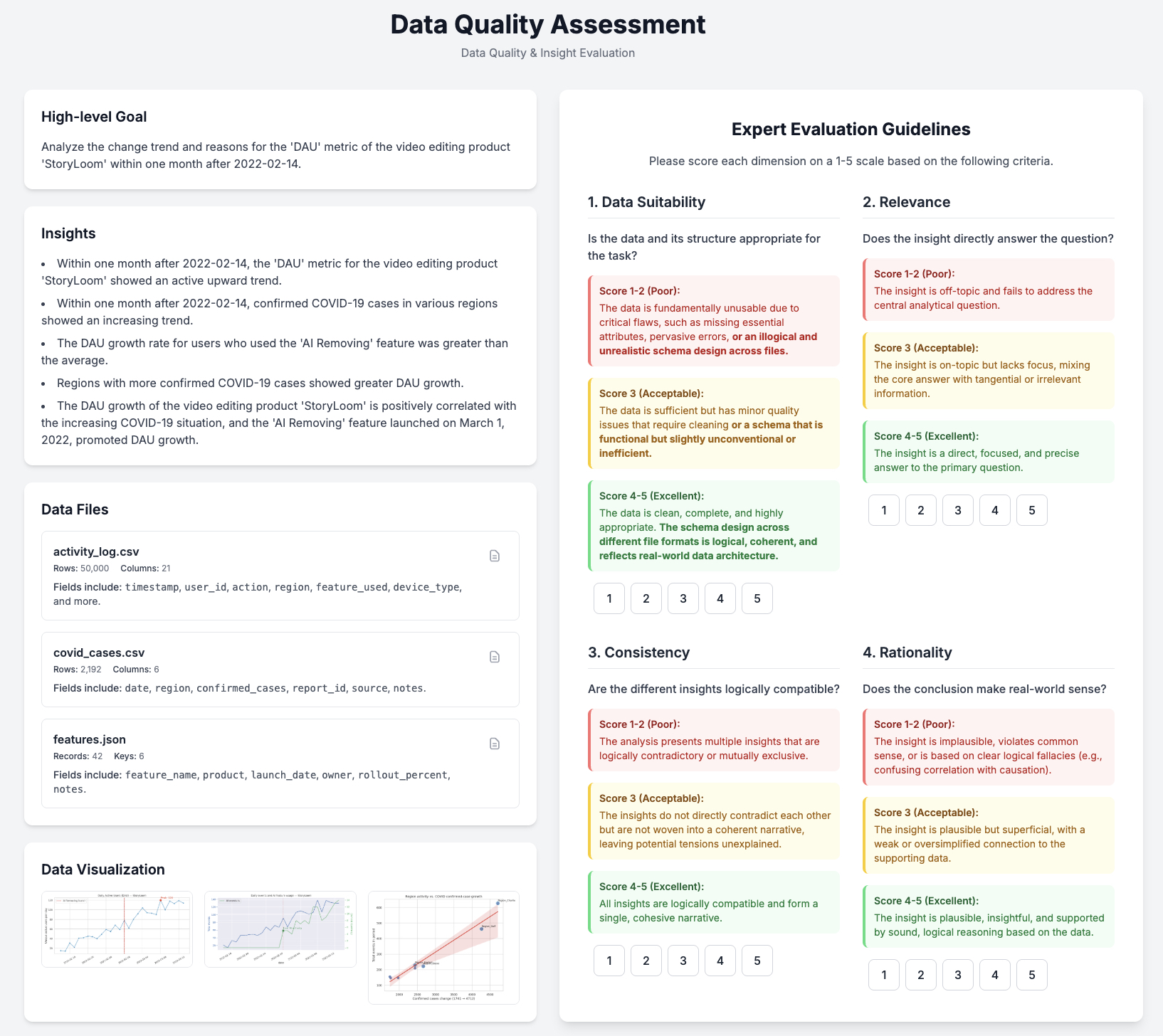} 
    \caption{Screenshot of the GUI for Benchmark Quality Assessment (Human Expert Scoring)}
    \label{fig:evaluation_gui} 
\end{figure*}

To validate the quality and robustness of UniDataBench, we conducted a rigorous evaluation from multiple perspectives. We employed both Large Language Models (LLMs) and human experts to assess the dataset, ensuring a comprehensive and objective analysis. This process was facilitated through a specifically developed GUI, as shown in Figure~\ref{fig:evaluation_gui}, which guided the expert reviewers through a structured review of the ground-truth included in the benchmark.

\subsection{Evaluation Metrics}
Our quality assessment is centered on four key dimensions that reflect the benchmark's core characteristics. These metrics are designed to be direct and clearly address the fundamental qualities of a data analysis task.

\begin{itemize}
[leftmargin=*,itemsep=2pt,topsep=0pt,parsep=0pt]

\item \textbf{Data Suitability:} Assesses the overall fitness of the provided data by evaluating both its \textbf{content quality} and \textbf{structural integrity}. 
\textbf{Content quality} involves verifying the presence of necessary attributes, appropriate data granularity, and the absence of critical flaws (e.g., an age of -1, excessive missing values). 
\textbf{Structural integrity} evaluates the realism and logical coherence of the schema design, particularly how different data formats (e.g., CSV for tabular, JSON for nested) are utilized to plausibly represent the dataset.

\item \textbf{Relevance:} Assesses whether the resulting insight is a direct and pertinent answer to the task's primary question. The focus is on ensuring all conclusions contribute purposefully to the analytical objective, rather than being sidetracked by tangential discoveries.

\item \textbf{Consistency:} Evaluates the logical coherence of the analysis. This ensures that insights derived from diverse data sources (e.g., CSV, DB, NoSQL, TXT) are free from internal contradictions and are not mutually exclusive.

\item \textbf{Rationality:} Assesses the real-world plausibility of the derived insights. This verifies that each conclusion is factually grounded and could manifest in reality, rather than being an artifact of flawed reasoning or representing an impossible scenario.
\end{itemize}

\begin{figure*}[t!]
    \centering
    \begin{subfigure}[b]{0.48\textwidth}
        \includegraphics[width=\linewidth]{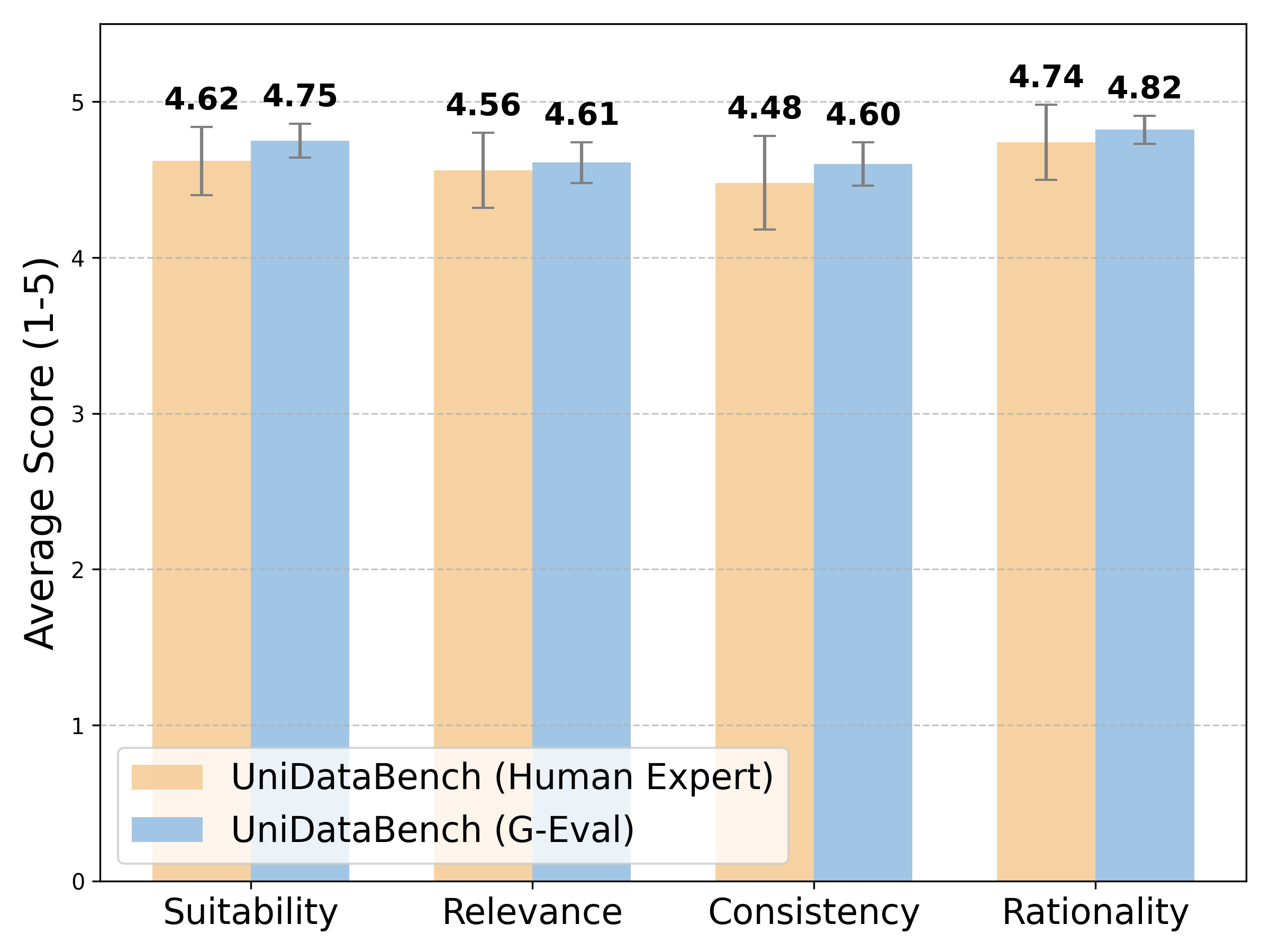}
        \caption{UniDataBench Quality Assessment.}
    \label{fig:quality_unidata}
    \end{subfigure}
    \hfill
    \begin{subfigure}[b]{0.48\textwidth}
        \includegraphics[width=\linewidth]{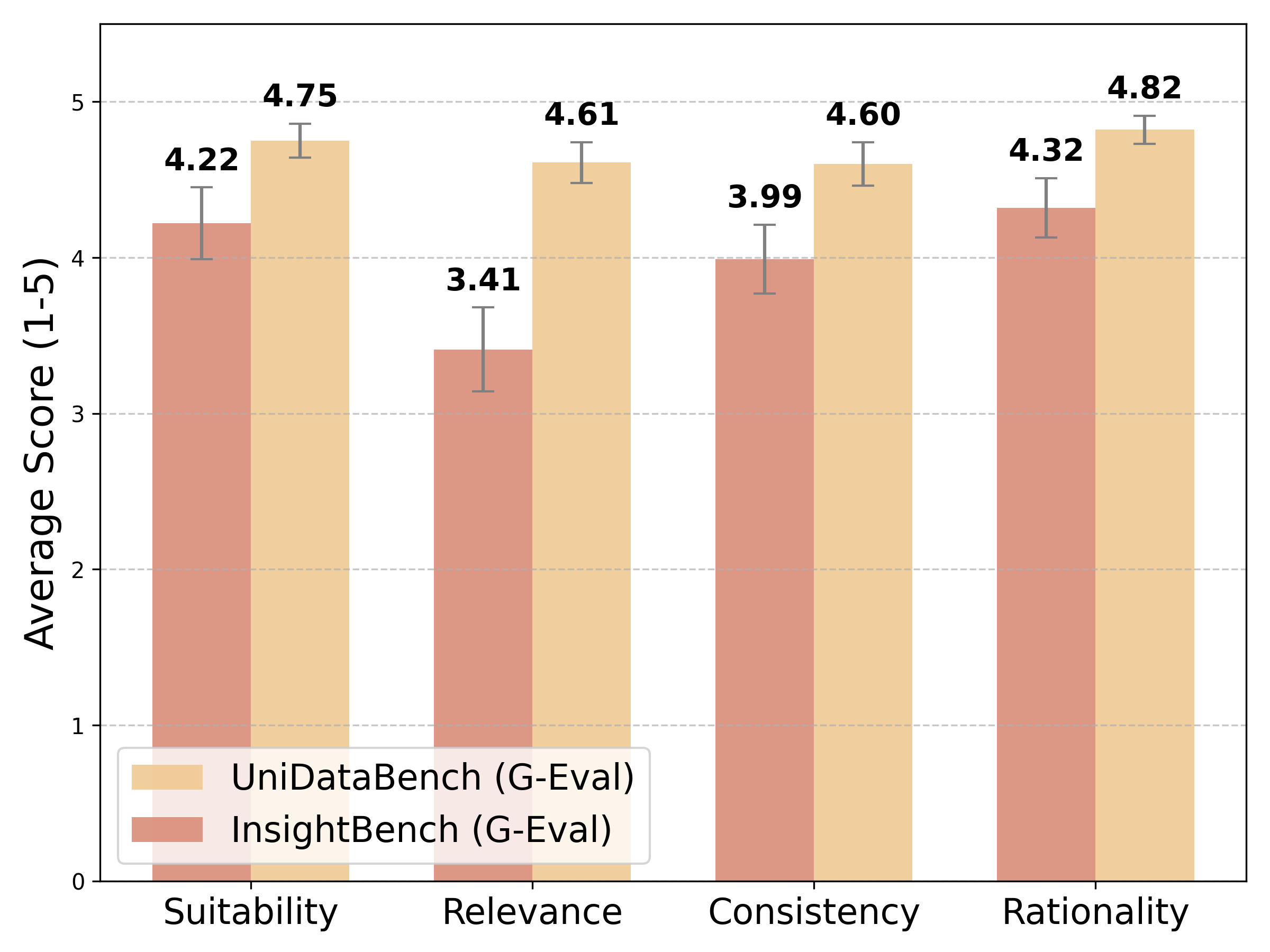}
        \caption{Comparative Analysis with InsightBench.}
    \label{fig:quality_comparison}
    \end{subfigure}
    \caption{Quality assessment of UniDataBench. (a) UniDataBench Quality Assessment comparing Human Expert and G-Eval evaluations. (b) Comparative Analysis of UniDataBench and InsightBench based on G-Eval.}
    \label{fig:quality_analysis}
\end{figure*}

\subsection{Evaluation Results}
We conducted a quality assessment of UniDataBench using evaluations from data analysis experts and the G-Eval \cite{liu-etal-2023-g} methodology. The same experts also evaluated tasks from InsightBench using our framework. As shown in Figure~\ref{fig:quality_analysis}, the evaluation confirms UniDataBench's high quality and its superior performance against InsightBench, particularly in task relevance.

\section{Detailed Experimental Setup}
\subsection{Baselines}
\label{appendix:baselines}

We benchmarked the following data analytics agents on UniDataBench: 
\begin{itemize}
[leftmargin=*,itemsep=2pt,topsep=0pt,parsep=0pt]
    \item \textbf{CodeGen}~\cite{majumder2024discoverybench}: CodeGen generates the entire analysis code in a single step based on the provided context, subsequently deriving insights and a workflow summary from the execution results. 
    \item \textbf{ReAct}~\cite{yao2023react}: ReAct solves tasks through an iterative process of multi-turn thought and action, where it incrementally generates and executes code to explore the data and uncover insights. 
    \item \textbf{Data-to-Dashboard}~\cite{zhang2025data}: Data-to-Dashboard is a modular multi-agent system that automates the generation of dashboards to produce insights supported by visualizations. 
    \item \textbf{Pandas Agent}~\cite{langchain2024pandas}: Pandas Agent is a LangChain-based agent, generates and executes Python code over a given data frame to produce direct answers to user queries as insights. 
     \item \textbf{DABstep Baseline Agent}~\cite{egg2025dabstepdataagentbenchmark}: The baseline agent introduced with the DABstep benchmark. It first inspects files in the working directory using a Python tool to plan its analysis path, then follows the ReAct paradigm to iteratively solve the task until a final answer is produced.
    \item \textbf{AgentPoirot}~\cite{sahu2024insightbench}: AgentPoirot is the baseline agent from InsightBench, which adopts a question-driven paradigm that first generates root questions and then formulates follow-up questions to delve deeper for insights. Since these agents natively support only CSV files, we adapted them to handle our multiple data sources for a comprehensive comparison.
    \item \textbf{ReActInsight}: ReActInsight is our proposed agent for analysing diverse data sources. It performs end-to-end analysis over diverse data formats by automatically discovering cross-source linkages, decomposing goals, and generating robust, self-correcting code to extract actionable insights.  
\end{itemize}

\subsection{Evaluation Metrics}
\label{appendix:metric}

Following other data analysis studies like Insight-Bench \cite{sahu2024insightbench}, we evaluate agent's performance using the G-Eval metric with summary-level and insight-level.

\begin{itemize}
[leftmargin=*,itemsep=2pt,topsep=0pt,parsep=0pt]
    \item \textbf{Summary-level evaluation}: we generate a comprehensive summary of insights provided by the agent. This summary is then compared against a ground-truth summary using the G-Eval score, which quantifies the overall coherence and correctness of the agent's output.

\item \textbf{Insight-level evaluation}: we conduct a more granular assessment. Let $G$ be the set of ground-truth insights and $pred\_insight$ be the set of insights generated by our agent. For each ground-truth insight $g \in G$, we identify the agent-generated insight $i \in I$ that maximizes the G-Eval similarity. The insight-level score is then calculated as the average of these maximum G-Eval scores across all ground-truth insights, as follows:

$$
score = \frac{1}{|G|} \sum_{g \in G} \max_{i \in I} \text{G-Eval}(g, pred\_insight) \quad
$$

\noindent where $|G|$ represents the total number of ground-truth insights, and $G(g, pred\_insight)$ denotes the G-Eval score between a ground-truth insight $g$ and an agent-generated insight $pred\_insight$.
\end{itemize}

This dual-level evaluation approach allows us to thoroughly assess both the broad understanding and the precise identification capabilities of our agent in complex data analysis tasks.

\section{More Experimental Results}

\subsection{Human Evaluation of ReActInsight}\label{appendix:human}

To supplement and validate the automatic G-Eval metrics, we also organized a small-scale yet rigorous human evaluation, the final results of which will be summarized in Table \ref{tab:human_eval}. During the task selection phase, we randomly sampled 30 questions from the complete UniDataBench test set, ensuring a balanced distribution of difficulty with 10 questions from each of the Easy, Medium, and Hard categories. We invited three senior data analytics practitioners to serve as independent evaluators, responsible for scoring the ``Insights'' and final ``Summary'' outputs generated by each agent. The evaluation process was structured around the following four core dimensions:
\begin{itemize}[leftmargin=*,itemsep=2pt,topsep=0pt,parsep=0pt]
    \item \textbf{Correctness}: Assessing the fidelity of numbers and facts in the generated content to the source data.
    \item \textbf{Insightfulness}: Evaluating whether the output goes beyond surface-level information to provide deeper perspectives and analysis.
    \item \textbf{Coherence \& Clarity}: Focusing on logical flow, readability, and the absence of internal contradictions.
    \item \textbf{Similarity to GT}: Measuring the semantic closeness of the generated content to the gold-standard answers provided by UniDataBench.
\end{itemize}

\begin{table}[h]
\centering
\resizebox{\linewidth}{!}{
\begin{tabular}{lccccccccccc}
\toprule
\multirow{2}{*}{\bf Evaluation Method} &
\multicolumn{5}{c}{\bf Insight\,–\,level} &
\multicolumn{5}{c}{\bf Summary\,–\,level} \\
\cmidrule(lr){2-6}\cmidrule(lr){7-11}
 & \bf D1 & \bf D2 & \bf D3 & \bf D4 & \bf Avg. &
   \bf D1 & \bf D2 & \bf D3 & \bf D4 & \bf Avg. \\ 
\midrule
Human Evaluation & 0.93 & 0.87 & 0.91 & 0.63 & 0.84 & 0.95 & 0.90 & 0.94 & 0.67 & 0.87 \\
LLM Evaluation   & 0.95 & 0.97 & 0.94 & 0.48 & 0.84 & 0.96 & 0.92 & 0.95 & 0.58 & 0.85 \\
\bottomrule
\end{tabular}}
\caption{Human \& LLM Evaluation: Individual Scores and Mean (0–1) for D1 Correctness, D2 Insightfulness, D3 Coherence and Clarity, and D4 Similarity to GT.}
\label{tab:human_eval}
\end{table}

\subsection{Hyperparameter Study}\label{appendix:hyper}
We conducted a hyperparameter study to assess the impact of the backbone LLM on performance, comparing gpt-4o, gpt-4-turbo, gpt-3.5-turbo, and llama-4-scout. As shown in Table~\ref{tab:agent_insight_scores}, the results confirm that the choice of LLM is a key factor. The gpt-4o variant consistently achieved the best results, securing the highest average scores in both Insight-level (0.4750) and Summary-level (0.5792) evaluations. While performance generally scaled with model capability, we noted that llama-4-scout (0.4618) was highly competitive, outperforming gpt-4-turbo (0.4547) on the average Insight-level score. Nevertheless, the dominant performance of gpt-4o, especially on medium and hard difficulty tasks, validates its selection as the most effective backbone for ReActInsight.

\begin{table*}[t!]
\centering
\resizebox{0.95\linewidth}{!}{
\begin{tabular}{lcccccccc}
\toprule
\multirow{2}{*}{\bf Agent} & \multicolumn{4}{c}{\bf Insight\,-\,level Scores (G-Eval)} & \multicolumn{4}{c}{\bf Summary\,-\,level Scores (G-Eval)} \\[-0.5mm]
\cmidrule(r){2-5}\cmidrule(r){6-9}
& \bf Easy & \bf Medium & \bf Hard & \bf Avg. & \bf Easy & \bf Medium & \bf Hard & \bf Avg. \\[0.5mm]
\midrule
ReActInsight (gpt-4o)  & \textbf{0.5030} & \textbf{0.4704} & \textbf{0.4517} & \textbf{0.4750} & 0.5984 & \textbf{0.5789} & \textbf{0.5605} & \textbf{0.5792} \\
ReActInsight (gpt-4-turbo)     & 0.4659 & 0.4688 & 0.4293 & 0.4547 & \textbf{0.6032} & 0.5441 & 0.5484 & 0.5652 \\
ReActInsight (gpt-3.5-turbo) & 0.4327 & 0.4561 & 0.4489 & 0.4459 & 0.5517 & 0.5361 & 0.5365 & 0.5414 \\
ReActInsight (llama-4-scout) & 0.4792 & 0.4708 & 0.4355 & 0.4618 & 0.5720 & 0.5386 & 0.5273 & 0.5460 \\
\bottomrule
\end{tabular}
}
\caption{Comparison of our agent performance using different LLMs on {\sc UniDataBench}.}
\label{tab:agent_insight_scores}
\end{table*}

\begin{figure}[ht!]
    \centering
    \includegraphics[width=0.99\linewidth]{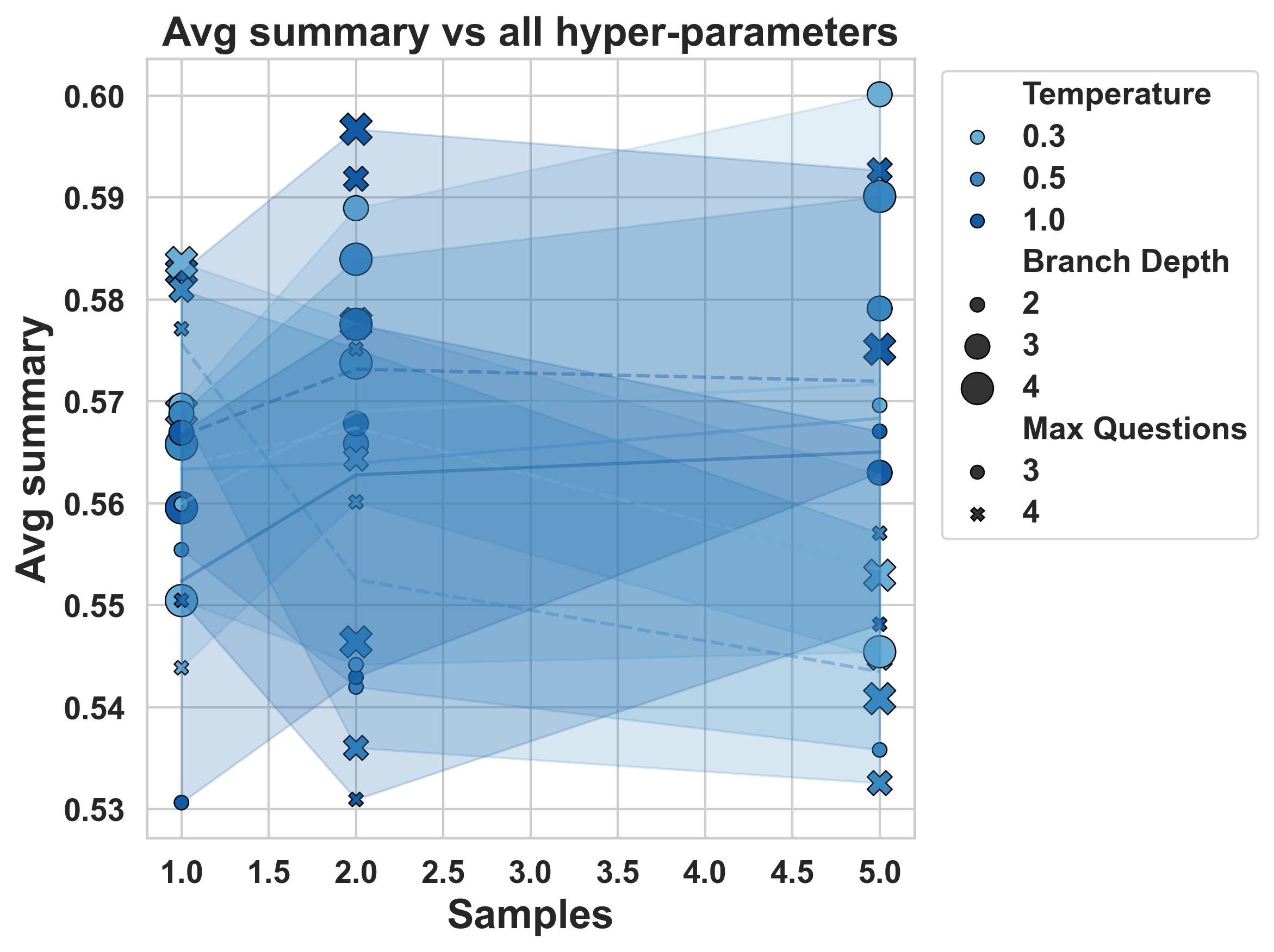}
    \vspace{-5pt}
    \caption{Hyperparameter Study for Summary.}
    \label{fig:hyper_sum}
    \vspace{-13pt}
\end{figure}

\subsection{Case Study}
\label{appendix:case_study}

\begin{table*}[t!]
\centering
\small
\begin{tabular}{p{4.6cm} p{4.6cm} p{4.6cm}}
\toprule
\textbf{UnidataBench Insights (GT)} & \textbf{AgentPoirot's Insights (Baseline)} & \textbf{ReActInsight's Insights (Ours)} \\
\midrule
\multicolumn{3}{c}{\textbf{Task Goal: \textit{Analyze the change in DAU metric for StoryLoom in Q2 2024}}} \\
\midrule
The increase in DAU in Q2 2024 is mainly attributed to \greenhl{the positive impact of the 'Fill Light Filter' feature}, while the 'Vintage Filter' feature had no significant effect on DAU growth.
&
\redhl{The lack of immediate DAU increase suggests that the new filters may require more time for user adoption or additional marketing efforts.}
&
\greenhl{The "Fill Light Filter" launched in Q2 2024 had the highest positive impact} on the 'DAU' metric for the 'StoryLoom' product, with total usage of 77,183 among intermediate account age users, 73,535 among new users, and 54,522 among long-term users.
\\
\midrule
\multicolumn{3}{c}{\textbf{Task Goal: \textit{Identify the peak sales quarter and peak profit month for the e-commerce 'MarketGrid' in 2018}}} \\
\midrule
The peak sales and profit periods are closely linked to targeted marketing activities: \greenhl{the 'Home\_Refresh' and 'Tech\_Launch' campaigns drove category-specific sales growth in Q1}, while the  \greenhl{'Winter\_Style' campaign in November was instrumental in maximizing profit through the Clothing category}. 
&
\redhl{November} 2018 stood out as the most profitable month for MarketGrid, achieving a total profit of \$270,000. This peak profit period \redhl{was influenced by targeted marketing campaigns such as "Tech\_Launch" and "Home\_Refresh."}
&
The top-performing marketing activities during \greenhl{the peak sales quarter were 'Tech\_Launch\_Mar' and 'Home\_Refresh\_Feb', contributing significantly to sales}. In contrast, \greenhl{the peak profit month (November 2018) was driven by 'Winter\_Style\_Launch'.}
\\
\midrule
\multicolumn{3}{c}{\textbf{Task Goal: \textit{Analysis of MoM Changes in the Number of Leads Converted to 'Vehicle Delivery' Created in August}}} \\
\midrule
The significant drop in conversion rate from \greenhl{"test drive" to "order" for the "VORTEX GT" model} is a key driver behind the overall decline in vehicle deliveries, suggesting targeted investigation and intervention for this model could improve outcomes.
&
The primary reason for this decline was a significant drop in the conversion rate from \greenhl{the 'test drive' to the 'order' stage} \redhl{[Note: here omit model analysis]}, which fell from 65\% to 52\% month-over-month across all models. Conversely, the final stage from 'Order' to 'Delivery' remained efficient, with a slight improvement in conversion rate, suggesting the bottleneck is clearly in the mid-funnel.
&
The primary driver for the overall decline in vehicle deliveries in August was a significant drop in the conversion rate from \greenhl{"test drive" to "order", specifically for the "VORTEX GT" model}, whose conversion rate at this stage plummeted from 45\% in July to 22\% in August.
\\
\midrule
\multicolumn{3}{c}{\textbf{Task Goal: \textit{Analyze the change in product order volume on the e-commerce platform 'ShopNext' between 2021 and 2020}}} \\
\midrule
The increase in order volume during these months is \greenhl{strongly correlated with a substantial rise in the number of promotions offered}, suggesting that promotional activities were a key driver.
&
Total revenue on ShopNext grew substantially by over 14\% from approximately \$70,989 in 2020 to \$81,161 in 2021, \redhl{supported by a notable surge in coupon usage} where orders using coupons more than doubled from 119 to 271.
&
\greenhl{Promotional activities in 2021, including increased counts} and diversified types like 'HolidayPromo' and 'MegaSale', led to a 40\% higher peak monthly order volume compared to 2020, with 140 orders in November and December in 2021 compared to 100 orders in December 2020.\\
\bottomrule 
\end{tabular}
\caption{
Comparative analysis of insights from ReActInsight and AgentPoirot. The baseline's failures (highlighted in red), such as factual inaccuracies and superficial conclusions, stem from its \textbf{inability to synthesize information across multiple data sources}. In contrast, ReActInsight (highlighted in green) leverages its multi-source exploration and linkage discovery mechanism to \textbf{effectively integrate heterogeneous data}. This capability enables it to produce quantitatively precise, granular, and robust analytical insights, demonstrating superior analytical depth.
}
\label{tab:case_study}
\end{table*}

To evaluate the quality of insights generated, we conduct case study against the strongest baseline, AgentPoirot, across four benchmark tasks, as shown in Table~\ref{tab:case_study}. ReActInsight demonstrates superior performance in two key areas.

\noindent\textbf{Factual Accuracy and Correct Attribution.}
Baseline models frequently generate factually incorrect or misattributed conclusions. For example, AgentPoirot incorrectly links peak profits to marketing campaigns that actually drove sales in a different quarter, demonstrating a confusion between correlation and causation. It also analyzes an irrelevant metric (revenue instead of order volume) in another task. ReActInsight avoids these pitfalls. Its \textbf{Hierarchical Planning} mechanism decomposes the main objective into logically coherent sub-questions, ensuring the analysis remains focused on the correct metrics and causal relationships, thus yielding factually accurate insights.

\noindent\textbf{Granular Root Cause Analysis.}
ReActInsight excels at identifying specific root causes, whereas the baseline often provides only superficial observations. In the vehicle delivery analysis, AgentPoirot notes a drop in a conversion rate but fails to identify the specific product model responsible. This failure arises from an inability to synthesize data across multiple sources. In contrast, ReActInsight's \textbf{Multi-Source Data Exploration} and \textbf{Entity-Graph} successfully link the sales funnel data with a separate product database. This allows it to pinpoint the issue precisely to the "VORTEX GT" model and quantify its conversion rate collapse from 45\% to 22\%. Similarly, it moves beyond a vague hypothesis in the StoryLoom analysis to identify a specific successful feature, the "Fill Light Filte", and provides detailed usage statistics across user segments, offering a far more actionable and granular insight.

\onecolumn

\section{Prompts}
\label{appendix:prompts}
\renewcommand\lstlistingname{Prompt} 
\begin{lstlisting}[breaklines=true,caption={Prompt for the agent in raising questions from multi sources data.}]
You are a senior data-analytics team lead.
Your job is to craft **answerable, high-impact questions** for the team,
given several heterogeneous datasets (csv/sql/nosql...).

Rules:
1. Only output what the user can copy-paste into their notebook.
2. NEVER add explanations outside the required <question> tags.
3. If the user sets question_type, bias the questions toward that type
   (e.g. descriptive / diagnostic / predictive / prescriptive).
   
We have SEVERAL datasets to analyse.  
<context>
{context}
</context>

Our business goal is:  
<goal>
{goal}
</goal>

Below is the merged schema and sample rows from **all data sources**:  
<schema>
{schema}
</schema>

You are asked to propose insightful questions for the team. These questions should be of a {question_type} nature.

Guidelines:
* Ask questions that help reach the stated goal.
* Consider relationships **across** datasets (joins, correlations, etc.).
* Each question must be realistic given ONLY the columns shown above.
* Do NOT number the questions.
* Generate at most {max_questions} questions and STOP.
* Wrap every question inside <question>...</question> tags.
### Response
\end{lstlisting}

\begin{lstlisting}[breaklines=true,caption={Prompt for the agent in raising followup questions from multi sources data.}]
You the manager of a data science team whose goal is to help stakeholders within your company extract actionable insights from their data.
You have access to a team of highly skilled data scientists that can answer complex questions about the data.
You call the shots and they do the work.
Your ultimate deliverable is a report that summarizes the findings and makes hypothesis for any trend or anomaly that was found.

Hi, I require the services of your team to help me reach my goal.

<context>{context}</context>

<goal>{goal}</goal>

<schema>{schema}</schema>

<question>{question}</question>

<answer>{answer}</answer>

Instructions:
* Produce a list of follow up questions explore my data and reach my goal.
* Note that we have already answered <question> and have the answer at <answer>, do not include a question similar to the one above. 
* Explore diverse aspects of the data, and ask questions that are relevant to my goal.
* You must ask the right questions to surface anything interesting (trends, anomalies, etc.)
* Make sure these can realistically be answered based on the data schema.
* The insights that your team will extract will be used to generate a report.
* Each question that you produce must be enclosed in <question>content</question> tags.
* Each question should only have one part, that is a single '?' at the end which only require a single answer.
* Do not number the questions.
* You can produce at most {max_questions} questions.
\end{lstlisting}

\begin{lstlisting}[breaklines=true,caption={Prompt for the agent in coding from multi sources data.}]
You are a meticulous senior data-scientist.
When the user asks for code, you must output **only ONE executable Python code block** that fulfils all requirements no extra commentary, markdown,
or explanation.  
If something would fail at run-time, handle it inside the code (raise a
clear Exception).  Be concise yet complete.


### Instruction

Business goal:
<goal>
{goal}
</goal>

Question to answer:
<question>
{question}
</question>

You have access to SEVERAL datasets whose schemas & samples follow:

<schema>
{schema}
</schema>

Raw file paths (csv / sqlite / json ...):

<database_paths>
{database_paths}
</database_paths>

Coding requirements:
1. Load whichever datasets / tables are needed to solve the question.  
   (pandas for csv/json, sqlite3+pandas for sqlite, json module for jsonl  ...)
2. Perform the analysis that directly answers the question.
3. Create **one or more** plots (use seaborn / matplotlib).  
    Name them sequentially:  plot_1.jpg, plot_2.jpg, ... 
4. For **each** plot i save four companion files in the working directory:  
       stat_i.json      key statistics used in your answer  
       x_axis_i.json    =< 50 representative x-values  
       y_axis_i.json    =< 50 representative y-values  
       plot_i.jpg       the image itself (already saved in step 3)  
   Every json must contain keys  "name", "description", "value".  
7. Make the script deterministic; set random seeds where randomness appears.
8. All files are saved under the {output_dir} directory

Output rules (IMPORTANT):
* Return exactly ONE code block that starts with ```python and ends with ```.
* Do NOT generate code blocks for other languages.
* No text outside the single code block.
* The code must execute all logic immediately at the top level; do not just define functions.

### Response
\end{lstlisting}

\begin{lstlisting}[breaklines=true,caption={Prompt for the agent in joining from multi sources data.}]
### Instruction

Business goal:
<goal>
{goal}
</goal>

Question to answer:
<question>
{question}
</question>

You have access to SEVERAL datasets whose schemas & samples follow:

<schema>
{schema}
</schema>

Raw file paths (csv / sqlite / json xxx):

<database_paths>
{database_paths}
</database_paths>

You must strictly follow the join_hints to perform data merging before creating any plots:
<join_hints>
{join_hints}
</join_hints>

Coding requirements:
1. Load whichever datasets / tables are needed to solve the question.  
   (pandas for csv/json, sqlite3+pandas for sqlite, json module for jsonl)  
   If <require_multi_join>true</require_multi_join>, you **must** join > 2 tables using the hints above.  
2. Perform the analysis that directly answers the question.
3. Create **one or more** plots (use seaborn / matplotlib).  
   Name them sequentially:  plot_1.jpg, plot_2.jpg,xxx  
4. For **each** plot *i* save four companion files in the working directory:  
      stat_i.json     -- key statistics used in your answer  
      x_axis_i.json   -- =< 50 representative x-values  
      y_axis_i.json   -- =< 50 representative y-values  
      plot_i.jpg      -- the image itself (already saved in step 3)  
   Each json must contain keys  `"name"`, `"description"`, `"value"`.  
5. Make the script deterministic; set random seeds where randomness appears.
6. Save **all** files under the directory **{output_dir}**.

Additional merging and processing requirements:
1. Standardize key values before merging:
   - For date keys: `df["date"] = pd.to_datetime(df["date"], errors="coerce").dt.normalize()`.
   - For category keys (e.g., province): `df["province"] = df["province"].astype(str).strip().lower()`.
   - Apply similar normalization (strip + lower) for other categorical keys like city or country.
2. Use a single merge result variable:
   - Use a consistent variable name (e.g., `merged`) for the merged result throughout the script. All statistics, plotting, and JSON exports must be based on this variable to avoid confusion.
3. Handle column name conflicts automatically:
   - When merging tables with overlapping column names, detect and resolve suffixes (e.g., `_x`, `_y`) using regex or column relationships.
   - Standardize critical columns (e.g., `confirmed_cases` or `dau`) by renaming them back to their original names.
4. Ensure the merge result is non-empty and provide fallback:
   - After an inner join, ensure the result is non-empty using `assert not merged.empty` or equivalent checks.
   - If empty, log a clear warning and automatically downgrade to an outer join to avoid data loss, while also providing diagnostics on key consistency.
5. Log key debugging information:
   - Log shapes of individual datasets and the merged result.
   - Log data types and sample values of keys used for joining.
   - For outer joins, optionally enable `indicator=True` to examine merge diagnostics.
6. Ensure safe paths and filenames:
   - Generate safe filenames using slugification and length limits: `safe = re.sub(r"[^-\w]", "_", text)[:60]`.
   - Use `pathlib.Path` for cross-platform path management.
7. Validate completeness of plots and JSON:
   - Ensure required columns for plotting are present and non-null before plotting. Use `pd.to_numeric(errors="coerce")` if needed and report missing ratios.
   - Write dates as strings in JSON (`dt.strftime("%Y-%m-%d)`).
   - Ensure consistent output structure for JSON files: `stat_x.json`, `x_axis_x.json`, `y_axis_x.json`. All fields must have clear meanings and non-empty values.
8. Limit resource usage and output sizes:
   - For images: Set `dpi =< 300`, long edge =< 4000 pixels, and file size =< 1 MB (adjust `dpi` or use `bbox_inches="tight"` if needed).
   - For JSON files: Limit entries and size (e.g., =< 10 KB). Use sampling or truncation if necessary.
9. Manage connections and file handles carefully:
   - Use `with` statements for database and file operations to ensure automatic resource cleanup.
   - Specify necessary parameters (e.g., `dtype`, `parse_dates`) when reading files to minimize type ambiguities.
10. Handle errors and provide diagnostic feedback:
    - Catch critical exceptions (e.g., `KeyError`, `ValueError`, `TypeError`, `pd.errors.ParserError`) and provide clear error messages with suggestions:
      - Missing columns: List available and expected column names.
      - Key mismatches: Indicate missing key values in each table.
      - Type issues: Recommend standardizing dates as `datetime` and categories as lowercase strings.
    - On failure, return a non-zero exit code or log `FAILED` to facilitate retries; include the latest exception summary in retry hints.

Optional but strongly recommended:
- Clearly label units and time ranges in legends, axis labels, and titles.
- Perform basic sanity checks on numerical columns (e.g., negative values or extreme outliers) and document handling methods (e.g., filling, truncation, or removal).
- Drop rows with NaN or use pairwise complete observations for correlation analyses; report the sample size used.

Output rules (IMPORTANT):
* Return exactly **one** code block that starts with ```python and ends with ```.
* Do **NOT** generate code blocks for other languages.
* No text outside the single code block.
* The code must execute all logic immediately at the top level; do not just define functions.
\end{lstlisting}

\begin{lstlisting}[breaklines=true,caption={Prompt for the agent in summarizing multi source data.}]
I require the services of your team to help me reach my goal.
<context>{context}</context>
<goal>{goal}</goal>
<history>{history}</history>
Instructions:
* Given the context, goal, and the full list of <question_i><answer_i> history pairs above, generate the 3 most important insights.
* Present the insights as a numbered list (e.g., "1. ... 2. ... 3. ..."), with each point focusing on a key, concise conclusion in plain language.
* Summarize highlights and findings; do not provide action recommendations.
* Make each insight as quantitative as possible, aggregating the findings.
* Enclose each insight within the <insight></insight> tag.
\end{lstlisting}

\end{document}